\documentclass[%
preprintnumbers,
 amsmath,amssymb,
 aps,
prfluids,
]{revtex4-2}

\usepackage{setspace}
\usepackage{graphicx}
\usepackage{amssymb}
\usepackage{epstopdf}
\usepackage{amsmath}
\usepackage{xcolor} 
\usepackage{ulem}
\usepackage{enumerate}

\newcommand{\pderiv}[2]{\frac{\partial #2}{\partial #1} }

\newcommand{\dz}{\ \hbox{d}z}

\newcommand{\Rey}{\hbox{Re}}
\newcommand{\Pran}{\hbox{Pr}}
\newcommand{\Ri}{\hbox{Ri}_b}
\newcommand{\R}{R_0}
\newcommand{\offset}{a_0}
\newcommand{\Rt}{R}
\newcommand{\at}{\mathbf{a}}

\newcommand{\Havg}[1]{\overline {#1}}
\newcommand{\Vavg}[1]{<{#1}>}

\newcommand{\ZOB}{z_{0,b}}
\newcommand{\ZOU}{z_{0,u}}

\newcommand{\TKE}{\hbox{TKE}}

\begin{document}
\title{The Dynamics of Asymmetric Stratified Shear Instabilities}
\author{Jason Olsthoorn}
\affiliation{Department of Civil Engineering, Queen's University, Kingston, Ontario, Canada, K7L 3N6}
\author{ Alexis K.~Kaminski}
\affiliation{Department of Mechanical Engineering, University of California
Berkeley, Berkeley, California, United States, 94720}
\author{Daniel M.~Robb}
\affiliation{Department of Civil Engineering, University of British Columbia, Vancouver, British Columbia, Canada, V6T 1Z4}

\begin{abstract}
Most idealized studies of stratified shear instabilities assume that the shear interface and the buoyancy interface are coincident. We discuss the role of asymmetry on the evolution of shear instabilities. Using linear stability theory and direct numerical simulations, we show that asymmetric shear instabilities exhibit features of both Holmboe and Kelvin-Helmholtz (KH) instabilities, and develop a framework to determine whether the instabilities are more Holmboe-like or more KH-like. Further, the asymmetric instabilities produce asymmetric mixing that exhibits features of both overturning and scouring flows and that tends to realign the shear and buoyancy interfaces. In all but the symmetric KH simulations, we observe a collapse in the distribution of gradient Richardson number ($Ri_g$), suggesting that asymmetry reduces the parameter dependence of KH-driven mixing events. The observed dependence of the turbulent dynamics on small-scale details of the shear and stratification has important implications for the interpretation of oceanographic data. 

\end{abstract}
\maketitle

\section{Introduction} \label{sec:intro}

Stratified turbulent mixing is a key physical process driving vertical transport in oceanic flows, setting the distribution of heat, salt, momentum, and other biogeochemical tracers. Because this mixing occurs on the smallest scales of fluid motion, it needs to be parameterized in large-scale circulation models. Furthermore, directly measuring the characteristics of oceanic turbulence is challenging, both due to the scales of the relevant motion and the sparseness of observations; parameterizations are thus also necessary for the interpretation of observational data.

Stratified mixing events are frequently modelled in terms of idealized shear instabilities.  While commonly discussed in the context of internal wave breaking in the stably-stratified ocean interior \citep{Garrett1972,Thorpe2018}, such instabilities have been observed in a wide variety of ocean contexts, including nonlinear internal waves near the continental shelf \citep{Moum2003}, shear at the base of the mixed layer \citep{Kaminski2021}, in the equatorial undercurrent \citep{Moum2011}, and flow over sills \citep{vanHaren2014}, as well as in estuaries \citep{Tedford2009,Holleman2016,Tu2020}. As a result, there is a  need to understand the dynamics of stratified shear instabilities in terms of resolved or measurable variables in order to accurately parameterize their effects \citep{Gregg2018}. 

One of the simplest examples of a stratified shear flow is that of a single sheared buoyancy interface. In this configuration, there are two limiting instabilities that may arise. When the shear is sufficiently strong relative to the stratification (typical of cases in which the velocity and buoyancy profile vary over a similar vertical scale), the flow is susceptible to Kelvin-Helmholtz (KH) instability. Conversely, where the stratification is sufficiently strong relative to the shear and the buoyancy varies over a smaller scale than the velocity (i.e.~the stratification is sharper than the shear), an alternate instability known as the Holmboe wave instability may arise \citep{Holmboe1962,Hazel1972}. 

Mixing events driven by KH instability are typically characterized by the formation of large overturns or ``billows'', which themselves become susceptible to secondary instabilities, the precise details of which depend on the flow parameters \citep{Corcos1976,Klaassen1991,Mashayek2012a,Mashayek2012b,Salehipour2015}, the background flow \citep{Lewin2021,VanDine2021}, and the initial conditions \citep{Brucker2007,Kaminski2019,Dong2019}. These secondary instabilities in turn trigger a transition to turbulence, leading to a brief period of vigorous mixing, which acts to smear out the initial velocity and buoyancy gradients (i.e.~``overturning'' turbulence as described by \citet{Woods2010,Caulfield2021}). After this period of intense mixing, the turbulent motions decay and the flow eventually relaminarizes. 

In contrast, Holmboe-driven mixing events (associated with sharp buoyancy interfaces) follow a different flow evolution. At finite amplitude, the initial linear instability leads to the formation of a pair of counterpropagating vortices on either side of the buoyancy interface (unlike the single large billow associated with KH). These vortices also support secondary instabilities \citep{Smyth1991,Smyth2003,Salehipour2016b}, which trigger a breakdown into turbulence. However, the turbulence is highest on either side of the interface, rather than in the centre of the shear layer. As a result, the turbulence ``scours'' the interface, preserving a sharp buoyancy gradient. While the initial instability growth is typically much slower than KH flows, the associated turbulence is longer-lived: this type of mixing event ``burns'', rather than ``flares'' \citep{Caulfield2021}. The evolution can depend on the underlying flow parameters, particularly the Reynolds number \citep{Salehipour2016b}; however, \citet{Salehipour2018} shows evidence of self-organized criticality for these flows, in which Holmboe instabilities with different initial stratifications evolve towards turbulent states with a common distribution of the gradient Richardson number (i.e.~the ratio of stratification and squared shear) featuring a peak near $1/4$. Overall, Holmboe-driven (scouring) mixing events can differ from the KH (overturning) paradigm, even for the same bulk flow parameters. 

Complicating the picture further, most previous studies of shear instabilities have only considered the configuration with symmetric shear and stratification about the buoyancy interface. However, asymmetric profiles frequently arise in many geophysical flows, whether due to flow geometry, flow history, or the forcing (e.g.~\citealp{vanHaren2014,Tu2020}). For example, the freshwater outflow in some estuaries is sufficiently strong to compete with the salt water intrusion of the ocean waters. In these cases, a stratified shear interface will form between the top freshwater outflow and the bottom salt water inflow. Both laboratory and field studies have shown that these stratified shear interfaces can have asymmetry \citep{Tedford2009,Yang2022}. It is therefore natural to ask how asymmetry affects the resulting nonlinear evolution and mixing in stratified shear flows. 

One simple way to introduce asymmetry into the background flow profile is to consider a sheared buoyancy interface where the shear and buoyancy profiles are vertically offset. \citet{Lawrence1991}, and later \citet{Carpenter2010} and \citet{Carpenter2011}, show that for asymmetric Holmboe-like profiles, the resulting linear normal-mode instability shares characteristics of both KH and Holmboe instability. In contrast to the symmetric case, they found a smooth transition between the different flow behaviours as characterized by the relative contributions of the kinematic and baroclinic fields.  


Asymmetric shear instabilities of this form have been observed both experimentally and numerically. In a series of laboratory experiments of shear layers offset from a thinner buoyancy stratification, \citet{Lawrence1991} showed that an offset shear layer led to the formation of a one-sided flow characterized by cusped waves (with stronger stratification) and asymmetric billows (with weaker stratification) that entrained wisps of fluid across the buoyancy interface. Similar features have been observed in subsequent experimental studies in which shear is driven above a buoyancy interface \citep{Strang2001}, in spatially-accelerating shear layers \citep{Pawlak1998,Yang2022}, and in sheared multilayer flows \citep{Caulfield1995}.

Similarly, in a direct numerical simulation study, \citet{Carpenter2007} showed that the nonlinear evolution of asymmetric Holmboe instabilities share characteristics of both KH- and Holmboe-like flow evolution: ``billow-like'' structures are observed to form that ``scour'' the interface. That is, the nonlinear evolution shows a mixture of behaviours, depending on the degree of asymmetry in the background flow profiles. Similar to the laboratory studies described above, the flows exhibited one-sided overturning features and cusped structures, and the asymmetric flows in general kept the initial interfaces intact. Further, \citet{Carpenter2007} showed that more asymmetry led to instabilities that were increasingly dominated by shear-layer vorticity (consistent with KH-like behaviour). The corresponding mixing depended non-monotonically on the degree of asymmetry, influenced by stronger three-dimensional motions as well as distance from the buoyancy interface. 

While the study by \citet{Carpenter2007} gave valuable insight into the impacts of asymmetry on a given mixing event, the simulations were limited to a single set of Reynolds, Richardson, and Prandtl numbers (defined in section~2), and the asymmetric cases were all Holmboe-like (with sharper buoyancy interfaces). Perhaps most significantly, due to the relatively low Reynolds number used, the resulting flow evolution was not necessarily turbulent; as the authors state, `\textit{the term ‘turbulence’ is being used loosely }[\ldots] \textit{to indicate a region of complex or chaotic flow structure, and the low Re used may preclude this flow from fitting descriptions of turbulent mixing in other studies}.' Consequently, there remain many open questions about turbulent mixing events in asymmetric stratified shear flows, motivating the present study.
We build upon the previous work on asymmetric shear instabilities by considering additional values of the Reynolds and Richardson numbers (and, crucially, a higher Reynolds number to explore turbulent effects), and by investigating symmetric and asymmetric profiles corresponding to both KH- and Holmboe-like setups.

The remainder of the paper will proceed as follows. In section~\ref{sec:linstab}, we introduce the linear instability problem and quantify the resulting normal-mode instabilities using the pseudomomentum approach described recently by \citet{Eaves2019}. Then, in section~\ref{sec:nonlin_sim}, we present the results of a series of direct numerical simulations corresponding to several symmetric and asymmetric cases. Consistent with the linear predictions, we find that the nonlinear evolution of the asymmetric cases exhibits a mixture of both KH and Holmboe behaviours, including both large billow-like overturning structures and regions of scouring. This combination of behaviours is both qualitative and quantitative: not only are the large-scale structures reminiscent of both pure KH and Holmboe instabilities, but the energetics and turbulent mixing parameters also exhibit both behaviours. As a result, even small amounts of asymmetry can lead to very different final velocity and buoyancy profiles when compared with the symmetric cases. For the same large-scale buoyancy and velocity jumps, a given mixing event can therefore smear out gradients, maintain sharp gradients, or some combination of these behaviours. That is, the turbulent dynamics depend sensitively on the small-scale details of the initial flow. Finally, in section~\ref{sec:conclusion} we conclude and put our results into context with recent work on stratified shear instabilities.


\section{Linear stability analysis} \label{sec:linstab}

\begin{figure}
    \centering
        \includegraphics[width=0.5\textwidth]{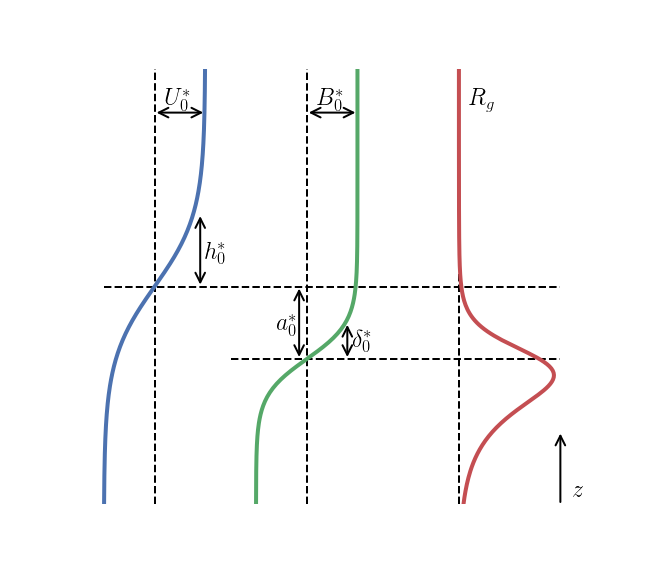}
    \caption{Representative profiles of the background horizontal velocity and buoyancy fields. The associated gradient Richardson number $Ri_g$ \eqref{eqn::RIG} for these profiles is also included. These profiles are similar to the asymmetric Holmboe case. }
    \label{fig:bgflow}
\end{figure}

We consider a vertically-asymmetric hyperbolic tangent stratified shear layer, 
\begin{gather}
    U^\ast(z^\ast)=U_0^\ast \tanh\left(\frac{z^\ast-\offset^\ast}{h_0^\ast}\right),
    \qquad \text{and} \qquad 
    B^\ast(z^\ast)=B_0^\ast \tanh\left(\frac{z^\ast}{\delta_0^\ast}\right) \, , \label{eqn:bgflow_dim}
\end{gather}
 which has the same form as \citet{Carpenter2007} and are shown schematically in figure~\ref{fig:bgflow}. The quantities $U_0^\ast$ and $B_0^\ast$ are half the horizontal velocity and buoyancy difference across the layers, respectively, where buoyancy is defined as $B^\ast=-g^\ast(\rho^\ast-\rho_0^\ast)/\rho_0^\ast$, with $\rho_0^\ast$ a reference density and $g^\ast$ the gravitational acceleration. The shear and buoyancy interfaces have initial half-widths of $h_0^\ast$ and $\delta_0^\ast$, respectively, and the centres of the two interfaces are vertically offset by $\offset^\ast$. Asterisks denote dimensional quantities. Nondimensionalizing velocity by $U_0^\ast$, buoyancy by $B_0^\ast$, and depth by $h_0^\ast$ gives 
\begin{equation}
    U(z)=\tanh \left(z-\offset\right)  \quad \text{and} \quad B(z)=\tanh\left(\R z \right) \, , \label{eqn:bgflow_nondim}
\end{equation}
where $\R\equiv h_0^\ast/\delta_0^\ast$ is the ratio of the initial interface thicknesses, and $\offset \equiv \offset^\ast /h_0^\ast $ is the initial profile offset.

The behaviour of two-dimensional linear perturbations to the background state $U(z)$, $B(z)$ is governed by the linearized Navier-Stokes equations under the Boussinesq approximation, written as
\begin{eqnarray}
    \frac{\partial \mathbf{u}}{\partial t} + \mathbf{u}\cdot\nabla\mathbf{U} + \mathbf{U}\cdot\nabla\mathbf{u} & = & -\nabla p + \Ri b \hat{\mathbf{k}} + \frac{1}{\Rey}\nabla^2\mathbf{u} \, , \label{eqn:NSlinear} \\
    \frac{\partial b}{\partial t} + \mathbf{u}\cdot\nabla B + \mathbf{U}\cdot\nabla b & = & \frac{1}{\Rey\ \Pran} \nabla^2 b \, , \label{eqn:buoyconvlinear} \\
    \nabla \cdot \mathbf{u} & = & 0 \, . \label{eqn:massctylinear}
\end{eqnarray}
Here, boldfaced variables denote vector quantities and lowercase $\mathbf{u}$, $b$, and $p$ denote perturbations to the velocity, buoyancy, and pressure, respectively. In the above, time has been nondimensionalized by the advective timescale $h_0^\ast/U_0^\ast$. In addition to $\R$ and $\offset$, the perturbation dynamics are governed by the Reynolds, Prandtl, and bulk Richardson numbers
\begin{equation}
    \Rey=\frac{U_0^\ast h_0^\ast}{\nu^\ast} \, , \qquad \Pran=\frac{\nu^\ast}{\kappa^\ast} \, , \qquad \Ri=\frac{B_0^\ast h_0^\ast}{U_0^{\ast 2}} \, , \label{eqn:dimensionlessnumbers}
\end{equation}
where $\nu^\ast$ is the kinematic viscosity and $\kappa^\ast$ is the diffusivity of the buoyancy field. We note here that $\Rey$ is defined based on the shear layer half-width and half-velocity difference, rather than the full shear layer thickness and velocity jump as in some parts of the literature, including \citet{Carpenter2007}.

\begin{figure*}
    \centering
    \includegraphics[width=0.8\textwidth]{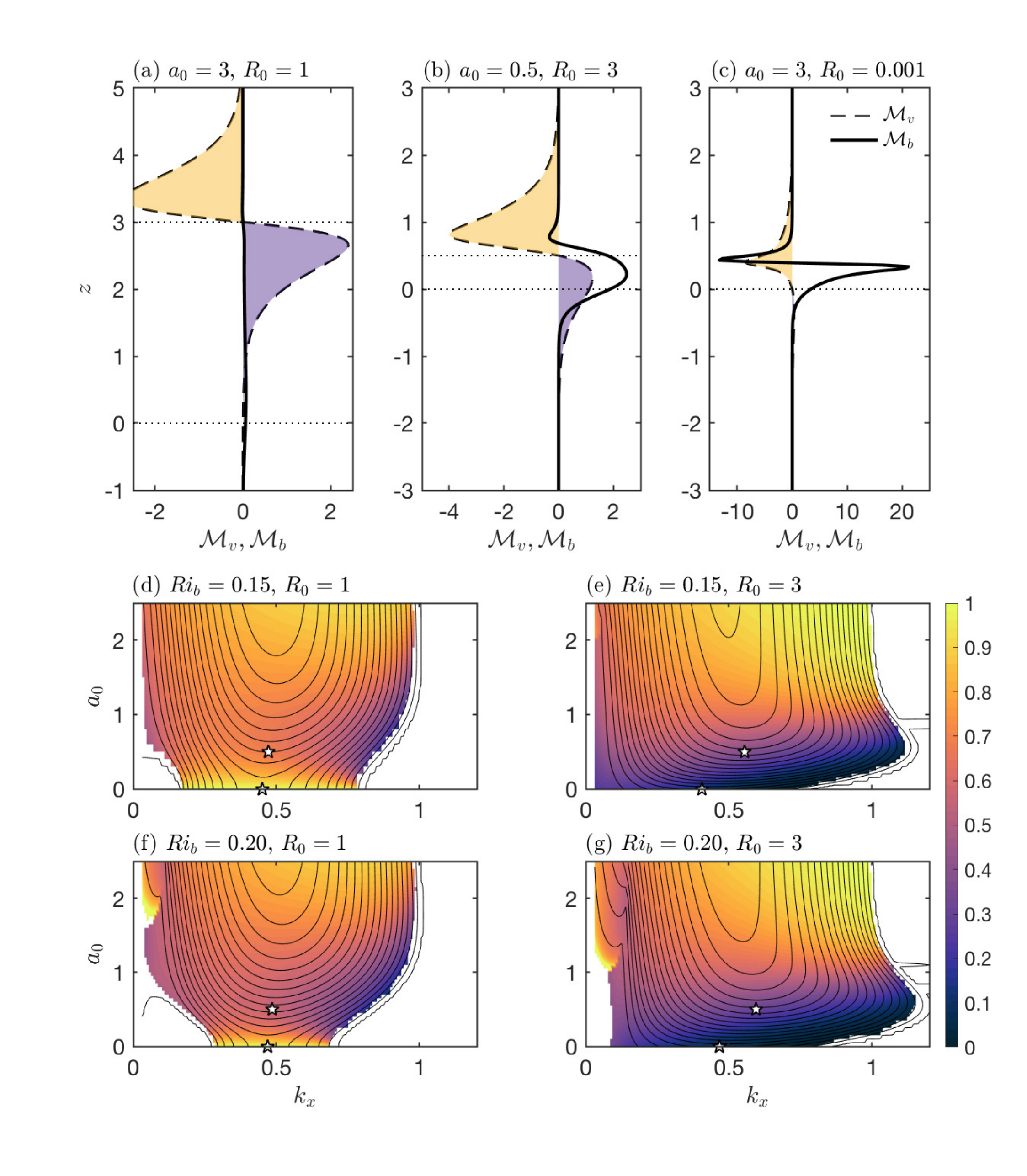}
    \caption{(Top row) Example profiles of $\mathcal{M}_v$ and $\mathcal{M}_b$ for the fastest-growing modes of (a) a KH-type instability far from from the interface ($\R=3$, $\offset=3$), (b) an asymmetric KH instability ($\R=1$, $\offset=0.5$), and (c) a near-symmetric Holmboe instability ($\R=3$, $\offset=0.001$). The horizontal dotted lines indicate the height of maximum shear and stratification, and the yellow and purple shaded regions represent contributions to $\mathcal{M}_v^-$ and $\mathcal{M}_v^+$, respectively. $\Ri=0.15$ for all three cases. (Middle and bottom rows) Growth rates (contours) and pseudomomentum ratios (colours) for the base flow~(\ref{eqn:bgflow_nondim}) for the symmetric and asymmetric KH and Holmboe configurations with $\Rey=1200$ and $\Pran=9$. (d) $\R=1$, $\Ri=0.15$. (e) $\R=3$, $\Ri=0.15$. (f) $\R=1$, $\Ri=0.20$. (g) $\R=3$, $\Ri=0.20$. The contour interval is 0.01 and the stars denote parameters for the nonlinear simulations described in table~\ref{Table:NumParams}.}
    \label{fig:sigma_pseudomtm}
\end{figure*}

To assess the linear stability of the shear layer~(\ref{eqn:bgflow_nondim}), we take the standard approach of assuming a normal-mode structure for the perturbations, 
\begin{equation}
    \left[\psi(x,z,t),b(x,z,t),p(x,z,t)\right]=\left[\hat{\psi}(z),\hat{b}(z),\hat{p}(z)\right]\mathrm{e}^{\mathrm{i}k_x(x-ct)} \, , \label{eqn:normalmodeansatz}
\end{equation}
where $\psi$ is the velocity streamfunction, $k_x$ the horizontal wavenumber and $c=c_r+\mathrm{i}c_i$ the complex phase speed of the perturbations. Substituting the normal-mode expressions into the governing equations (\ref{eqn:NSlinear})-(\ref{eqn:massctylinear}) and rearranging gives the Taylor-Goldstein equation including viscous and diffusive effects (implemented by \citet{Smyth2011}). Prescribing $k_x$, we solve for the corresponding eigenvalue $c$ and the associated eigenfunctions $\left[\hat{\psi}(z),\hat{b}(z),\hat{p}(z)\right]$; regions of linear instability correspond to modes with $c_i\ge0$, indicated by the contours in figure~\ref{fig:sigma_pseudomtm}.

Solutions of the Taylor-Goldstein equation predict which modes are unstable, their associated growth rate and phase speed, and the vertical structure of the eigenmodes. For the background flow~\eqref{eqn:bgflow_nondim}, we also want to predict the \textit{nature} of the linear instability: are the growing modes more like a KH instability or a Holmboe instability? 

To answer this question, we recall that stratified shear instabilities can be described in terms of interacting waves. Stratified parallel shear flows can support two different types of waves (namely vorticity waves, associated with changes in the shear, and internal gravity waves, associated with the stratification). The different linear growth mechanisms can be described in terms of interactions between different combinations of these waves \citep{Carpenter2011}. Within this framework, the KH instability arises from a resonant interaction between two vorticity waves, while the Holmboe instability arises from the interaction between a vorticity wave and an internal gravity wave. In this sense, the Holmboe instability is an inherently stratified instability \citep{Alexakis2009,Carpenter2011,Caulfield2021}. (A third named linear instability, the Taylor-Caulfield instability, can be thought of as the result of two internal waves interacting \citep{Caulfield1995,Carpenter2011,Eaves2019}. As the stratification considered here has only a single buoyancy interface and therefore only supports a single internal gravity wave, it does not support the Taylor-Caulfield instability.)

Using the wave-interaction description of the different instability mechanisms, the qualitative nature of a given unstable mode can be predicted using the vertical structure of the eigenfunction. For example, \citet{Carpenter2010} used a method of ``partial growth rates'' to quantify the contribution to the overall growth rate from buoyancy and vorticity interfaces, allowing for the description of modes in terms of KH- and Holmboe-like behaviour. More recently, \citet{Eaves2019} used a pseudomomentum-based approach to classify linear modes, again in terms of contributions from the vorticity and the stratification, allowing for a description of modes as being more or less like KH, Holmboe, or Taylor-Caulfield instabilities. 

Here, we apply the pseudomomentum-based approach of \citet{Eaves2019} to classify the linear modes arising in the asymmetric stratified shear layer~(\ref{eqn:bgflow_nondim}). In our notation, Eaves and Balmforth defined
\begin{equation}
\mathcal{M}_v=\frac{1}{2}\frac{U''\hat{\psi}^2}{\left| U-c \right|^2} 
\quad \text{and} \quad
\mathcal{M}_b=\frac{-\Ri (U-c_r)B'\hat{\psi}^2}{\left|U-c\right|^4} \label{eqn:M_b}
\end{equation}
as the contributions to the pseudomomentum from the background vorticity and stratification, respectively, 
where the primes denote derivatives with respect to $z$. 
It can be shown that a requirement for exponentially-growing modes is that $M=\langle \mathcal{M}_v+\mathcal{M}_b\rangle=0$, where angle brackets denote the integral over the spatial domain \citep{Eaves2019}. 

This requirement that $M=0$ can be used to characterize different modes in terms of the different canonical instabilities. To do so, we define the ratio
\begin{equation}
\mathcal{R}_M=\frac{\langle \mathcal{M}_v H( \mathcal{M}_v)\rangle}{\langle-\mathcal{M}_v H(-\mathcal{M}_v)\rangle}=\frac{M_v^+}{M_v^-} \, , \label{eqn:instab_ratio}
\end{equation} 
where $H$ is the Heaviside function and $M_v^+$ and $M_v^-$ are defined as the magnitudes of the contributions of positive and negative $\mathcal{M}_v$, respectively, to $M$ 
(i.e.~$\langle \mathcal{M}_v \rangle =M_v^+-M_v^-$). 
This ratio can be used to describe the character of the asymmetric instabilities considered here. In particular, we find two limiting cases of interest for flows described by~(\ref{eqn:bgflow_nondim}). In the first case, if a given unstable mode is associated purely with interactions from vorticity with no buoyancy contribution ($\mathcal{M}_b=0$ as is expected for a pure KH mode), then the requirement that $M=0$ for a growing mode implies $\langle \mathcal{M}_v \rangle=0$, and so $\mathcal{R}_M=1$ in the KH limit. In the second case, if the positive contributions from vorticity are balanced entirely by negative contributions from the stratification, 
then $M_v^-=\langle \mathcal{M}_b \rangle$ 
and $M_v^+=0$, giving $\mathcal{R}_M=0$ (as expected for a pure Holmboe mode). We note that the definition of $\mathcal{R}_M$ in (\ref{eqn:instab_ratio}) reflects our choice to introduce asymmetry in the mean profiles by shifting the shear interface above the buoyancy interface. As a result, the leftward-propagating Holmboe wave above the buoyancy interface is amplified relative to the rightward-propagating wave below the interface, i.e.~the upper Holmboe wave is the fastest-growing mode for these asymmetric profiles. For a flow in which the shear interface is offset below the buoyancy interface, or one in which the lower Holmboe wave is isolated, $1/\mathcal{R}_M$ would give the same classification results as described below. For the profiles considered in~(\ref{eqn:bgflow_nondim}) with $\offset>0$, $0\le\mathcal{R}_M\le1$. 

We illustrate these limiting cases in the top row of figure~\ref{fig:sigma_pseudomtm}. For a case in which the shear layer is significantly offset from the stratified layer (figure~\ref{fig:sigma_pseudomtm}a), for which the most unstable mode is very similar to an unstratified KH mode, the positive (purple) and negative (yellow) vorticity contributions cancel almost perfectly, with only a small buoyancy contribution. On the other hand, for a near-symmetric Holmboe mode (figure~\ref{fig:sigma_pseudomtm}c), the negative contributions from vorticity are almost entirely cancelled out by the net positive buoyancy contribution, with only a very small positive vorticity contribution. Finally, for some modes the negative contribution from vorticity is balanced by both a positive vorticity contribution and a buoyancy contribution (figure~\ref{fig:sigma_pseudomtm}b), suggesting that the instability shares characteristics of both a KH and a Holmboe mode.

We then use the pseudomomentum ratio $\mathcal{R}_M$ to classify the predicted unstable modes of (\ref{eqn:bgflow_nondim}) across a range of wavenumbers $k_x$ and vertical offsets $\offset$ for different $\Ri$ and $\R$, as shown in figure~\ref{fig:sigma_pseudomtm}(d-g). For $R_0=3$, we find consistent results to those described in \citet{Carpenter2007}: increasing asymmetry shifts the linear instability from a Holmboe-like behaviour at small $\offset$ to a KH-like behaviour at large $\offset$, and a mixture of the two instabilities at intermediate $\offset$. For $R_0=1$, as expected we find KH-like modes for $\offset=0$ and for large $\offset$ (when the shear layer is far enough removed from the initial stratification to behave essentially like an unstratified shear layer). However, between these limits asymmetry in the background flow leads to modes with both KH-like and Holmboe-like behaviour, with $\mathcal{R}_M<1$ and nonzero phase speed (not shown); this effect is stronger for higher $\Ri$ (figure~\ref{fig:sigma_pseudomtm}f). That is, mixed modes are possible even when the background shear and stratification vary over similar vertical scales.



\section{Nonlinear simulations} \label{sec:nonlin_sim}

\subsection{Implementation}

Linear theory predicts that the shear instabilities associated with an asymmetric shear layer exhibit behaviours similar to both KH and Holmboe instabilities. What is the associated nonlinear behaviour of these instabilities? To answer this question, we perform a series of direct numerical simulations of the fully nonlinear Navier-Stokes equations under the Boussinesq approximation, 
\begin{eqnarray}
        \frac{\partial \mathbf{u}}{\partial t} + \mathbf{u}\cdot\nabla\mathbf{u} & = & -\nabla p + \hbox{Ri}_b b \hat{\mathbf{k}} + \frac{1}{\hbox{Re}}\nabla^2\mathbf{u} \, , \label{eqn:NS_u}\\
        \frac{\partial b}{\partial t} + \mathbf{u}\cdot\nabla b & = & \frac{1}{\hbox{Re Pr}} \nabla^2 b \, , \\
        \nabla \cdot \mathbf{u} & = & 0 \, .\label{eqn:NS_div}
\end{eqnarray}
The Reynolds number (\Rey), Prandtl number (\Pran) and Richardson number ($\Ri$) are identical to those defined above. Unlike equations~(\ref{eqn:NSlinear}-\ref{eqn:massctylinear}) in the previous section, $\mathbf{u}$ and $b$ are the \textit{total} (nonlinear) velocity and buoyancy fields.

The simulations were performed using the Spectral Parallel Incompressible Navier-Stokes Solver (SPINS) \citep{Subich}. SPINS implements pseudospectral spatial derivatives in all three directions and an explicit third-order time stepping scheme. The horizontal domain is periodic, and the vertical gradient-free boundaries are imposed using a cosine-transformation. The streamwise extent ($L_x$) of the computational domain is selected to fit one wavelength of the fastest growing linear instability (see \S\ref{sec:linstab}), and the vertical and spanwise extents are chosen to match those from \citet{Carpenter2007}.

The initial conditions consist of the velocity and buoyancy profiles~(\ref{eqn:bgflow_nondim}) perturbed with the eigenfunction of the fastest growing linear mode predicted from the Taylor-Goldstein equation ($\mathbf{u}_e, b_e$) and a random velocity perturbation in the form of normally-distributed random noise. The amplitude of the eigenfunctions and the random noise are $\varepsilon_e=0.01$ and $\varepsilon_N=0.001$, respectively. These initial amplitudes are comparable to several previous shear instability studies (e.g.~\citealp{Salehipour2016b,Dong2019,VanDine2021}).

Our simulations have four flow configurations, determined by the relative width $\R$ and offset $\offset$ of the initial velocity and buoyancy profiles: 
\begin{enumerate}
    \item SKH -- ``symmetric Kelvin-Helmholtz'' flow with $\R=1$ and $\offset=0$;
    \item SHI -- ``symmetric Holmboe'' with $\R=3$ and $\offset=0$;
    \item AKH -- ``asymmetric Kelvin-Helmholtz'' flow with $\R=1$ and $\offset=0.5$;
    \item AHI -- ``asymmetric Holmboe'' flow with $\R=3$ and $\offset=0.5$.
\end{enumerate} 
For each configuration, we consider three sets of Reynolds and Richardson numbers, for a total of twelve simulations as summarized in table~\ref{Table:NumParams}. These cases span values of $\mathcal{R}_M$ from $0$ to $1$, i.e.~Holmboe-like to KH-like linear instabilities. We note that the SKH, SHI, and AHI simulations with $\Rey=300$ and $\Ri=0.15$ use the same parameters considered by \citet{Carpenter2007}. All simulations were continued until the flow relaminarized and there was no significant overturning of the buoyancy field. Consistent with previous DNS studies of shear instability (e.g.~\citealp{Smyth2000,Salehipour2016b,Kaminski2019}), the grid spacing ($
\Delta x$) was selected to be less than three times the minimum Batchelor scale ($L_{B,\mathrm{min}}$), defined by the maximum horizontally averaged dissipation rate $\Havg{\varepsilon}$, 
\begin{gather}
    L_{B,\mathrm{min}}= \left( \frac{1}{\hbox{Re}^3 \max \Havg{\varepsilon}}\right)^{\frac 14} \left(\frac{1}{\hbox{Pr}}\right)^\frac 12 \, ,
\end{gather}
where the overbar $\Havg{\cdot}$ denotes the horizontal mean. 

Our computational setup and flow parameters have been chosen to prioritize setting $Pr=9$ (a realistic value for heat in water). To do so has required some compromise in both the streamwise extent of the domain and the choice of Reynolds number. We revisit this point in the conclusions.

\begin{table*}
\centering
\begin{tabular}{cc|cc|ccccc|c|c}
Case & Instability &  Domain Size  & Resolution & $\Rey$ & $\Ri$ & $\Pran$ & $\R$  & $\offset$ &  $\frac{\Delta x}{ L_B}$ & $\mathcal{R}_M$ \\
& &  ($L_x \times L_y \times L_z$) & ($N_x \times N_y \times N_z$) &&&&&\\
\hline
1 & SKH & $14.02 \times 9 \times 18$ & $256 \times 192 \times 384$ & 300 & 0.15 & 9 & 1 & 0.0 & 2.8 & 1.00 \\ 
2 & AKH & $13.45 \times 9 \times 18$ & $256 \times 192 \times 384$ & 300 & 0.15 & 9& 1 & 0.5 & 2.7 & 0.66 \\
3 & AHI & $11.47 \times 9 \times 18$ & $256 \times 192 \times 384$ & 300 & 0.15 & 9& 3 & 0.5 & 2.4 & 0.37 \\
4 & SHI & $16.52 \times 9 \times 18$ & $384 \times 192 \times 384$ & 300 & 0.15 & 9& 3 & 0.0 & 2.4 & 0.09 \\
5 & SKH & $13.94 \times 9 \times 18$ & $768 \times 512 \times 1024$ & 1200 & 0.15 & 9& 1 & 0.0 &  2.5 & 1.00 \\
6 & AKH&  $13.30 \times 9 \times 18$ & $768 \times 512 \times 1024$& 1200 & 0.15 & 9& 1 & 0.5 & 2.4 & 0.68 \\ 
7 & AHI & $11.32 \times 9 \times 18$ & $768 \times 512 \times 1024$ & 1200 & 0.15 & 9& 3 & 0.5 & 2.8 & 0.39 \\
8 & SHI & $15.50 \times 9 \times 18$ & $768 \times 512 \times 1024$ & 1200 & 0.15 & 9& 3 & 0.0 & 2.9 & 0.11 \\
9 & SKH & $13.35 \times 9 \times 18$ & $512 \times 384 \times 768$ & 1200 & 0.20 & 9& 1 & 0.0 & 2.9 & 1.00 \\ 
10 & AKH & $12.96 \times 9 \times 18$ & $768 \times 512 \times 1024$ & 1200 & 0.20 & 9& 1 & 0.5 & 2.4 & 0.56 \\
11 & AHI & $10.58 \times 9 \times 18$ & $768 \times 512 \times 1024$ & 1200 & 0.20 & 9& 3 & 0.5 & 2.5 & 0.30 \\
12 & SHI & $13.42 \times 9 \times 18$ & $768 \times 512 \times 1024$ & 1200 & 0.20 & 9& 3 & 0.0 & 2.5 & 0.07
\end{tabular}
\caption{Summary of parameters for the suite of direct numerical simulations. We use the following naming convention: (i) SKH -- symmetric Kelvin-Helmholtz instability, (ii) AKH -- asymmetric Kelvin-Helmholtz instability, (iii) AHI -- asymmetric Holmboe instability, (iv) SHI -- symmetric Holmboe instability, corresponding to the definitions in the main text.}
\label{Table:NumParams}
\end{table*}

\subsection{Flow evolution} \label{sec:flowevol}

\begin{figure*}
    \centering
    \includegraphics[width=0.9\textwidth]{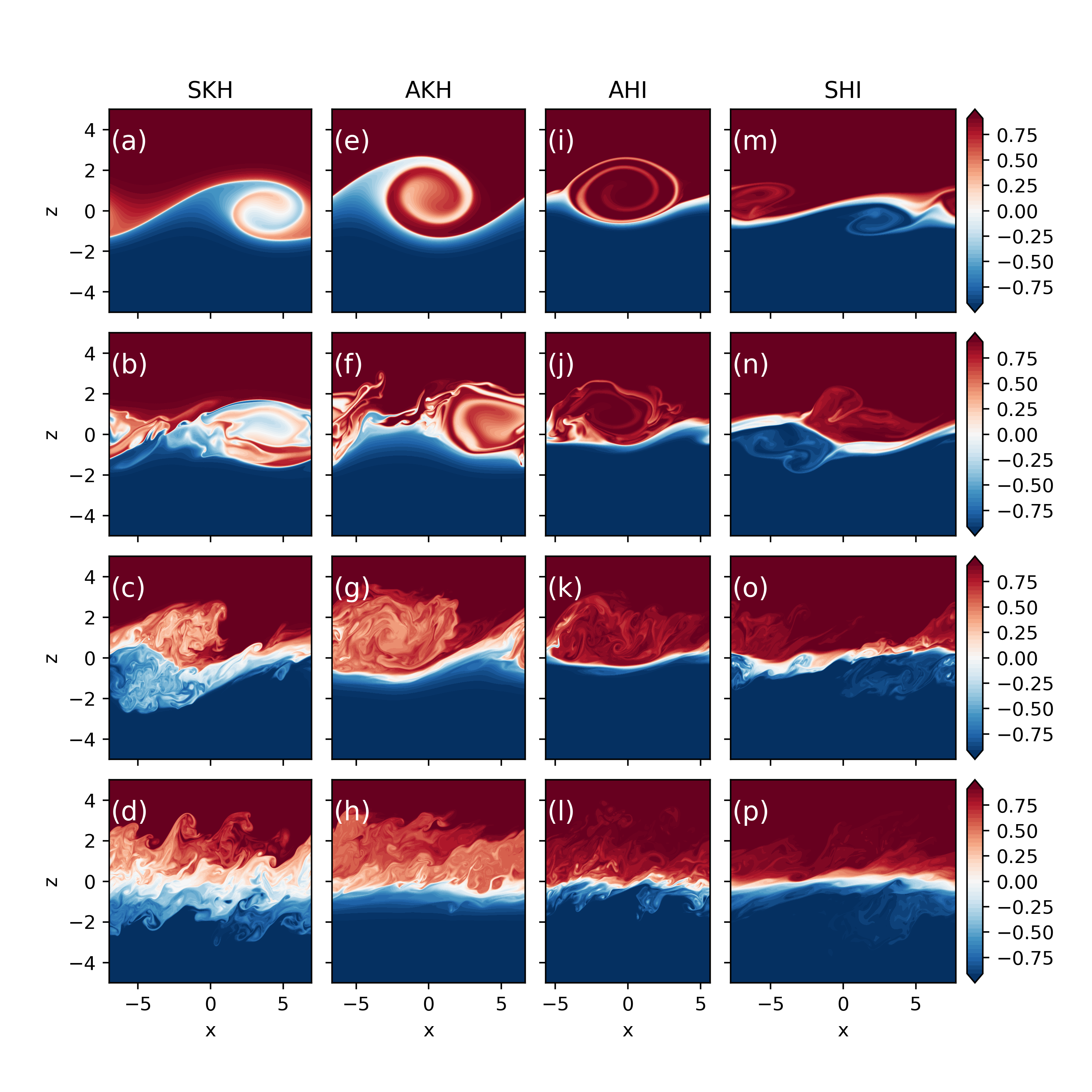}
    \caption{Snapshots of buoyancy field at representative times throughout the flow evolution. Vertical ($x-z$) slices are provided for (a)-(d) Case 4 -- $\R=1$, \ $\offset=0, \ t=\{55,85,115,145\}$, (e)-(h) Case 5 -- $\R=1$, $\offset=0, \ t=\{55,85,115,145\}$, (i)-(l) Case 6 -- $\R=\frac 13$, $\offset=0, \ t=\{70,105,140,175\}$, and (m)-(p) Case 6 -- $\R=\frac 13$, $\offset=0, \ t=\{120,170,220,270\}$. In all included cases $\Rey=1200,\  \Ri=0.15,\ \Pran=9.$}
    \label{fig:SliceEvolution}
\end{figure*}

We have highlighted in \S~\ref{sec:linstab} that the linear instabilities of the offset profiles~\eqref{eqn:bgflow_nondim} exhibit a continuum in structure between the pure KH and the pure Holmboe instabilities. Here, we will show that the mixed features of the linear instabilities result in nonlinear flow features that are reminiscent of both KH and Holmboe mixing events. 

\begin{figure}
    \centering
    \includegraphics[width=0.6\textwidth]{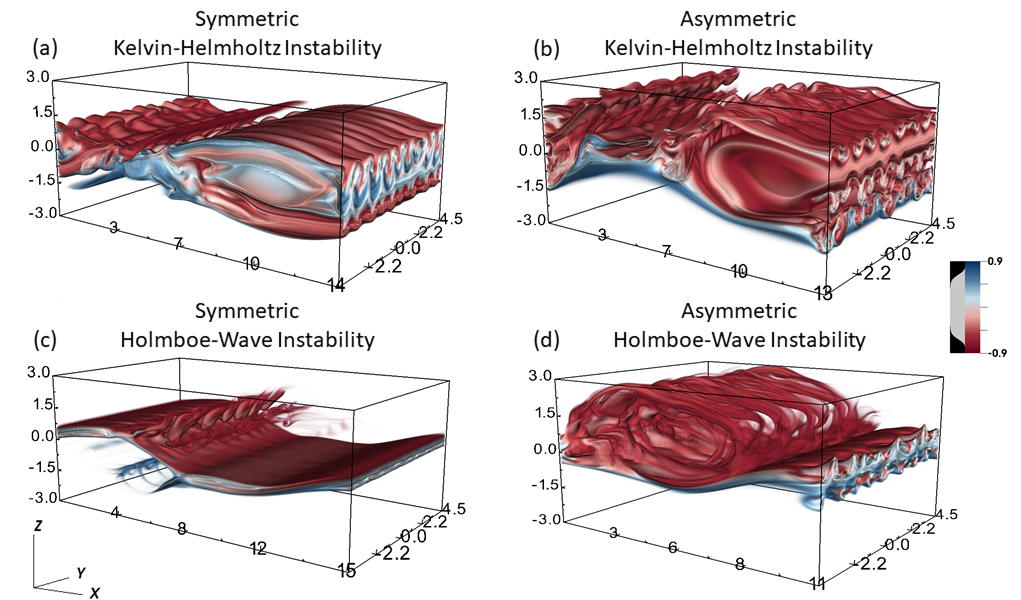}
    \caption{Three-dimensional visualizations of the buoyancy field as the secondary instabilities develop. The output times are $t=85$ (SKH), $t=85$ (AKH), $t=105$ (AHI), and $t=170$ (SHI),~i.e. the same times as in figure~\ref{fig:SliceEvolution}(b,f,j,n). These plots were constructed using VisIt's \citep{VisIt} Volume plot, which sets the opacity of the buoyancy field based on its value.}
    \label{fig:3DFig}
\end{figure}

 Illustrating the flow evolution for each of the four cases (SKH, AKH, AHI, and SHI), figure~\ref{fig:SliceEvolution} shows vertical ($x-z$) slices of the buoyancy field at representative times. We focus on the simulations with $\Rey=1200$ and $\Ri=0.15$, and will note significant differences for the other $\Rey$ and $\Ri$ cases. We complement these two-dimensional slices with plots of the three-dimensional structures of the buoyancy field (figure~\ref{fig:3DFig}) at the same times as in  figure~\ref{fig:SliceEvolution}(b,f,j,n), illustrating the onset of secondary instabilities in each case. We discuss the key features of each case below:  
\begin{enumerate}[(i)]
    \item \textbf{SKH}: The nonlinear evolution of the symmetric Kelvin-Helmholtz instability (SKH) is illustrated in figure~\ref{fig:SliceEvolution}(a-d). The flow follows the typical evolution of a KH-driven mixing event \citep{Caulfield2000}. The initial linear instability leads to the formation of the classic billow structure at finite amplitude, overturning the entire stratified interface (figure~\ref{fig:SliceEvolution}a). This billow is stationary with respect to the mean flow and approximately vertically symmetric. The billow then becomes unstable to three-dimensional secondary instabilities (figure~\ref{fig:SliceEvolution}b and \ref{fig:3DFig}a), which trigger a transition to turbulence that fills the entire shear layer (figure~\ref{fig:SliceEvolution}c,d). The buoyancy profile is more diffuse at the end of the simulation, with nearly equal mixing of the top and bottom fluid layers. While the same sequence is observed in our SKH simulations at $\Rey=300$ and $\Ri=0.20$, the turbulent mixing in those cases is significantly weaker due to the stronger effects of viscosity and stratification, respectively. That is, the turbulent SKH dynamics are sensitive to the flow parameters for the simulations presented here.

    \item \textbf{SHI}: The evolution of the symmetric Holmboe instability (SHI) is shown in figure~\ref{fig:SliceEvolution}m-p. SHI is characterized by waves propagating along the buoyancy interface, leading to the formation of counterpropagating vortices at finite amplitude (figure~\ref{fig:SliceEvolution}m). For the background flow considered here, these vortices are symmetric about the buoyancy interface. Like the SKH billow, the vortices are themselves susceptible to three-dimensional secondary instabilities (figure~\ref{fig:SliceEvolution}n and \ref{fig:3DFig}c), which trigger a transition to a fully-turbulent flow (figure~\ref{fig:SliceEvolution}o). Unlike SKH, however, the resulting mixing is longer-lived and acts to scour the initial stratification, maintaining a sharp interface (figure~\ref{fig:SliceEvolution}p). We will return to this point in section~\ref{sec::MeanQuantities}. In addition, the SHI is more consistent across $\Rey$ and $\Ri$, suggesting that the overall turbulent dynamics may be less sensitive to the initial flow parameters.

    \item \textbf{AHI}: When the buoyancy profile is thinner than the shear profile, and a vertical offset is introduced (i.e.~$\R=3$ and $\offset=0.5$), the asymmetric Holmboe instability (AHI) may arise (figure~\ref{fig:SliceEvolution}i-l). As seen previously by \citet{Carpenter2007}, at finite amplitude the AHI leads to the formation of a large propagating vortex above the buoyancy interface that entrains part of the buoyancy interface to form a billow-like structure. At $\Rey=300$, this primary vortex produces a series of ejection events before decaying. However, at larger $\Rey=1200$, these ejections are interrupted by the rapid formation of secondary instabilities (figure~\ref{fig:3DFig}d). In either case, the resultant mixing is primarily above the buoyancy interface, leading to an asymmetric final buoyancy profile. Consequently, the turbulent flow only overturns part of the initial interface, maintaining a strong buoyancy gradient on the lower side.

     \item \textbf{AKH}: Finally, we consider the situation where the shear and stratification have the same initial thickness but are vertically offset (i.e.~$\R=1$ and $\offset=0.5$), giving rise to the asymmetric Kelvin-Helmholtz instability (AKH), depicted in figure~\ref{fig:SliceEvolution}(e-h). This instability results in the formation of a large billow, similar to that of the SKH. However, there are two key differences: the billow is centred above the buoyancy interface, and is no longer stationary with respect to the mean flow (as demonstrated by the different horizontal locations between panels in figure~\ref{fig:SliceEvolution}). The billow is then susceptible to secondary instabilities which are qualitatively similar to those observed for the SKH (figure~\ref{fig:3DFig}b) and likewise trigger a transition to turbulence. However, similar to the AHI flow, the mixing in the AKH case is vertically asymmetric: while the turbulent mixing acts to smear out the buoyancy gradient above the initial stratified interface, the lower part of the interface remains relatively sharp (figure~\ref{fig:SliceEvolution}h).

\end{enumerate}

Consistent with the linear predictions in section~\ref{sec:linstab}, the nonlinear evolution and mixing associated with the asymmetric cases (AKH and AHI) share qualitative features of both KH- and Holmboe-driven mixing events. For example, they exhibit both a billow structure and a non-zero phase speed. This trend is consistent across the different parameter ($\Rey$, $\Ri$) scenarios presented in table~\ref{Table:NumParams}, with the most notable difference being that at lower $\Rey$ and higher $\Ri$, the SKH is much less energetic. We explore this comparison more quantitatively in the following sections.

\subsection{TKE budget} \label{sec:TKE}

\begin{figure}
    \centering
    \includegraphics[width=0.9\textwidth]{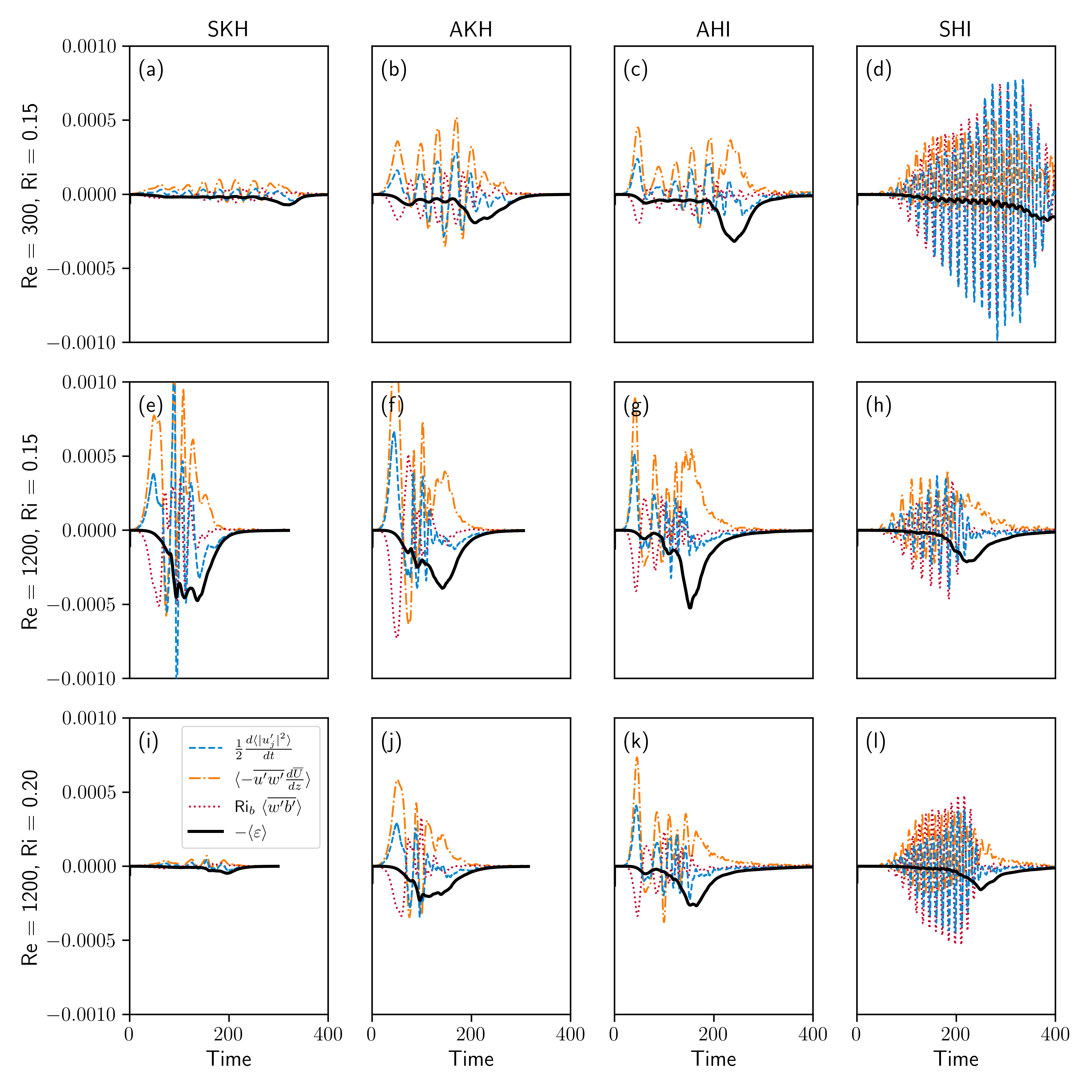}
    \caption{TKE budgets for the cases listed in table \ref{Table:NumParams}. Note that the time axes were trimmed (e.g. panel d) to have all of the TKE budgets on a consistent and legible horizontal scale.}
    \label{fig:TKEBudget}
\end{figure}

As we have seen, shear instabilities mix the stratification such that the final mean buoyancy and shear profiles are more diffuse than they were initially.
The evolution of the mean flow can be distinguished from that of the shear instabilities through a Reynolds decomposition. 
That is, 
\begin{gather}
    \mathbf{u} = \Havg{\mathbf{u}} + \mathbf{u}^\prime, \qquad b = \Havg{b} + b^\prime,
\end{gather}
where $\Havg{\ \cdot\ }$ denotes the horizontal average. We will use the fact that $\Havg{\mathbf{u}} \approx (\Havg{u}_1(z,t),0,0)^T$ below.
Under this decomposition, the volume-integrated turbulent kinetic energy ($\TKE =  \Vavg{\frac 12 |u_i^\prime|^2 }$)  evolves as
\begin{gather}
  \frac{d}{dt} \hbox{TKE} = \underbrace{\Vavg{-\Havg{u_i^\prime u_j^\prime} \pderiv{x_j}{\Havg{u_i}}}}_{\hbox{Production}} + 
  \underbrace{
  \vphantom{\frac{x_i}{\overline{U_j}}}
  \Ri \ \Vavg{\Havg{u_3^\prime b}}}_{\hbox{Buoyancy Flux}}  
  - \underbrace{
  \vphantom{\frac{x_i}{\overline{U_j}}}
  \Vavg{\varepsilon^\prime}}_{\hbox{Dissipation}}, \label{eqn:TKEBudget}
\end{gather} 
where $u_i$ is the $i^{th}$ component of $\mathbf{u}$, and the turbulent dissipation $\varepsilon^\prime = \frac 1{2\ \Rey} \left( \pderiv{x_i}{u^\prime_j} + \pderiv{x_j}{u^\prime_i}\right)\left( \pderiv{x_i}{u^\prime_j} + \pderiv{x_j}{u^\prime_i}\right) $. We are implicitly summing over repeated indices.

A physical interpretation of \eqref{eqn:TKEBudget} is that TKE is produced by extracting energy from the mean flow, and will decay via viscous dissipation and mixing (a negative buoyancy flux). Equation \eqref{eqn:TKEBudget} enables us to make a quantitative comparison between the cases listed in table \ref{Table:NumParams} (see figure~\ref{fig:TKEBudget} as well as the time-integrated flux, production, and dissipation values in table~\ref{Table:FlowProperties}), and provides another measure of how asymmetry frustrates the traditional distinction between the SKH and the SHI.  

We first focus on Case~5 from table~\ref{Table:NumParams}, i.e.~SKH with $\Rey=1200$ and $\Ri=0.15$ (figure~\ref{fig:TKEBudget}e). The growth of the initial billow is associated with a large shear production and negative buoyancy flux, as dense fluid is transported upwards. The billow quickly becomes unstable to secondary instabilities and the flow becomes turbulent, enhancing viscous dissipation. Eventually, the dissipation dampens the TKE, and the flow relaminarizes. The duration of the turbulent event is relatively short, with few oscillations between the TKE production and the buoyancy flux. We note that the SKH flows with lower $\Rey$ (figure~\ref{fig:TKEBudget}a) and higher $\Ri$ (figure~\ref{fig:TKEBudget}i) are much less energetic, owing to the stronger effects of viscosity and stratification, respectively.

On the other hand, for Case~8 (SHI at the same parameter values; figure~\ref{fig:TKEBudget}h), the buoyancy flux, shear production, and the rate of change of TKE exhibit a sequence of oscillations associated with periodic conversions between kinetic and potential energy in the propagating Holmboe waves. These oscillations continue until the onset of secondary instabilities and the subsequent transition to turbulence with increased $\varepsilon$. Finally, the flow relaminarizes with all terms decaying. In contrast to the SKH case, the SHI event is long-lived, lasting many wave periods. A similar behaviour is observed for the other parameter values (figure~\ref{fig:TKEBudget}d,l), though we note that the oscillations persist for a longer time at lower $\Rey$ as the flow remains laminar for a longer time.

Consistent with the linear predictions in \S~\ref{sec:linstab} and the qualitative descriptions in \S~\ref{sec:flowevol}, asymmetry in the base flow results in a hybrid of the pure KH (SKH) and pure Holmboe (SHI) behaviours. Figures~\ref{fig:TKEBudget}(f,g) demonstrate that as the instability mechanism transitions from KH-like to Holmboe-like, the duration of the turbulent mixing event increases. Both an initial period of positive shear production and negative buoyancy flux as well as regularly-spaced oscillations in the budget terms are observed, similar to the initial billow development in the SKH flow and the periodic waves in the SHI flow. While viscosity and stratification still impact the resulting flow evolution for the asymmetric cases (particularly for the AKH flows), we note that in contrast to the SKH flows, the asymmetric cases remain more energetic at lower $\Rey$ (figure~\ref{fig:TKEBudget}b,c) and higher $\Ri$ (figure~\ref{fig:TKEBudget}j,k).

Once again, asymmetry in the base flow results in a hybrid flow between the pure KH (SKH) and the pure Holmboe-wave instability (SHI). Figure \ref{fig:TKEBudget} demonstrates that as the instability mechanism transitions from KH-like to Holmboe-like, the duration of the turbulent event increases (e.g.~figure \ref{fig:TKEBudget}f) and regularly-spaced oscillations appear between the rate of change of TKE and buoyancy flux (e.g.~figure \ref{fig:TKEBudget}g). Thus, as discussed in \S \ref{sec:linstab}, asymmetry introduces an apparent transition between the SKH and the SHI.

The volume-averaged quantities provide an overview of the energetics of these symmetric and asymmetric flows. Further insight can be gleaned from considering {\it where} buoyancy flux or dissipation are occurring within the flow, which we consider in the next section.

\subsection{Horizontally-averaged quantities} \label{sec::MeanQuantities}

\begin{figure*}
    \centering
    \includegraphics[width=0.95\textwidth]{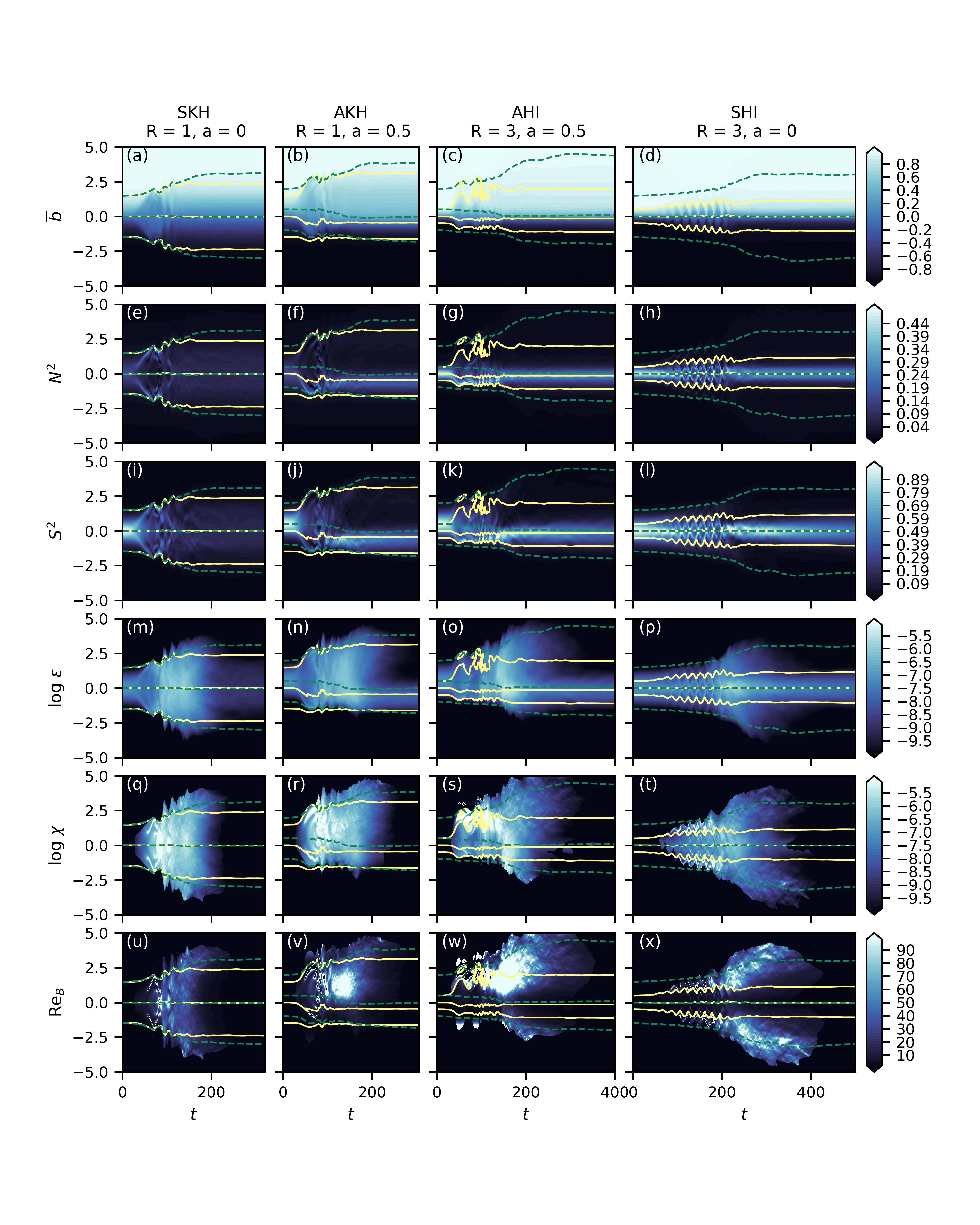}
    \caption{Contour plots of the horizontally-averaged (a)-(d)  buoyancy, (e)-(h) $N^2$, (i)-(l) $S^2$, (m)-(p) $\log \varepsilon$, (q)-(t) $\log \chi$, and (u)-(x) Re$_B$ as a function of time. Data is plotted for Case 4 (left), Case 5 (middle left), and Case 6 (middle right) and Case 7 (right) -- $\Rey=1200, \Ri=0.15$. Superimposed on each plot are the (-0.9,0,0.9) contours of $\overline{\textbf{u}}$ (green) and $\overline b$ (yellow). Anomalous values of large $\chi$ and Re$_B$ associated with $N^2\rightarrow0$ have been masked our as described in \S~\ref{sec:SelfOrg}.}
    \label{fig:MeanFields}
\end{figure*}

The TKE budgets above show that the turbulent flow field extracts energy from the mean flow, which is subsequently dissipated. We now discuss how that energy transfer changes the vertical structure of the horizontally-averaged mean flow. In particular, we characterize the structure of the mean velocity and buoyancy profiles through the buoyancy frequency ($N^2$) and shear rate ($S^2$), defined 
\begin{gather}
    N^2 = \pderiv{z}{\Havg{b}},\qquad S^2 = \left(\pderiv{z}{\Havg{u_1}}\right)^2.
\end{gather}
Similarly, we will present the mean viscous dissipation ($\varepsilon$) and scalar variance dissipation ($\chi$), computed as 
\begin{alignat}{2}
    \varepsilon &= \frac 1{2\Rey} \Havg{\left( \pderiv{x_i}{u_j} + \pderiv{x_j}{u_i}\right)\left( \pderiv{x_i}{u_j} + \pderiv{x_j}{u_i}\right)} &&\approx \frac{1}{\Rey} S^2  + \varepsilon^\prime,\qquad
    \\
    \chi &= \frac{1}{N^2} \frac{1}{\Rey\ \Pran }\overline{\pderiv{x_i}{b}\pderiv{x_i}{b}} &&\approx \frac{N^2}{\Rey \ \Pran}  + \chi^\prime.
\end{alignat}
High $\varepsilon$ and $\chi$ indicate regions of strong turbulent motions acting on the velocity and buoyancy fields, respectively. 

  The timeseries of horizontally-averaged buoyancy ($\bar{b}$), buoyancy frequency ($N^2$), shear rate ($S^2$), viscous dissipation ($\varepsilon$), and scalar variance dissipation ($\chi$) for $\Rey=1200$ and $\Ri=0.15$ are presented in figure \ref{fig:MeanFields}. Superimposed on each plot are the (-0.9, 0, 0.9) contours of $\Havg{u_1}$ (green) and $\Havg b$ (yellow), to approximate the bottom, middle and top of the buoyancy and velocity interfaces. 

As described above, the SKH is characterized by an initial billow structure that breaks down into turbulence, which subsequently decays. There is a corresponding ``burst'' of intense turbulence, which produces an associated increase in viscous and scalar dissipation (figure~\ref{fig:MeanFields}m,q). This turbulence rapidly mixes (thickens) the initial buoyancy and velocity interfaces, decreasing the overall stratification and shear (figure~\ref{fig:MeanFields}a,e,i). Consistent with previous studies of KH instability, the overall thicknesses of the buoyancy and velocity interfaces (as shown by the green and yellow contours) evolve similarly to one another throughout the flow evolution \citep{Smyth2000}.


In contrast, the wavelike nature of the SHI results in oscillations in the horizontally-averaged fields at the buoyancy interface (figure~\ref{fig:MeanFields}d). The maximum gradient of the buoyancy and velocity interfaces remains relatively large after the mixing event (figure~\ref{fig:MeanFields}h,l), and the buoyancy interface remains thinner than the velocity interface. We also observe that the viscous and scalar dissipation rates remain high for an extended period throughout the mixing event, consistent with the ``burning'' behaviour described in \citet{Caulfield2021}.

To recap, KH-driven mixing is associated with overturning turbulence, acting to smear out the initial interface. In contrast, Holmboe-driven mixing is associated with scouring turbulence, keeping interfaces sharp \citep{Woods2010,Caulfield2021}; these behaviours are well demonstrated by the evolution of $N^2$ and $S^2$ for the SKH and SHI simulations (figure~\ref{fig:MeanFields}e,h,i,l). However, asymmetry breaks this dichotomy. We find that the asymmetric instabilities both mix the buoyancy interface (identified by the yellow contours), while preserving a sharp buoyancy gradient that is offset from its initial position. This hybrid behaviour is a result of the asymmetric mixing of the AKH and AHI. 

The locations of the viscous dissipation and mixing are identified through $\Havg{\varepsilon}$ and $\Havg{\chi}$. For both the SKH (figure~\ref{fig:MeanFields}m,q) and SHI (figure~\ref{fig:MeanFields}q,t), $\varepsilon$ and $\chi$ are roughly symmetric with respect to the buoyancy interface. Conversely, AKH (figure~\ref{fig:MeanFields}n,r) and AHI (figure~\ref{fig:MeanFields}o,s) show more mixing above the buoyancy interface, associated with the breakdown of the initial billow structure. We find that the $\Havg{u_1}$ = [-0.9,0.9] contours provide reasonable vertical bounds for the location of the mixing and dissipation.
Thus, the initial asymmetry of both AHI and AKH results in asymmetric mixing of the buoyancy and velocity fields, \textit{while preserving the buoyancy and velocity interfaces}. As such, asymmetric shear instabilities combine the intense mixing of a KH instability with the interface-preserving property of a Holmboe instability.

We have argued that, in contrast to SKH, the AKH, AHI, and SHI exhibit scouring behaviour, which preserves the buoyancy interface. To illustrate this, we define the buoyancy Reynolds number ($Re_B = \varepsilon/\left(\nu N^2\right)$) as a measure of the regions where the turbulence is most intense relative to the background stratification. 
As describe qualitatively above, while the SKH has elevated values of $Re_B$ throughout the buoyancy interface (figure~\ref{fig:MeanFields}u), we find high values of $Re_B\approx 100$ for the AKH, AHI, and SHI away from the buoyancy interface (figure~\ref{fig:MeanFields}v-x). That is, the SKH overturns and diffuses the buoyancy interface, while the other cases scour but do not eliminate a strong buoyancy gradient.

\subsection{Interface position and width} \label{sec:InterfPos}

\begin{figure}
    \centering
    \includegraphics[width=0.75\textwidth]{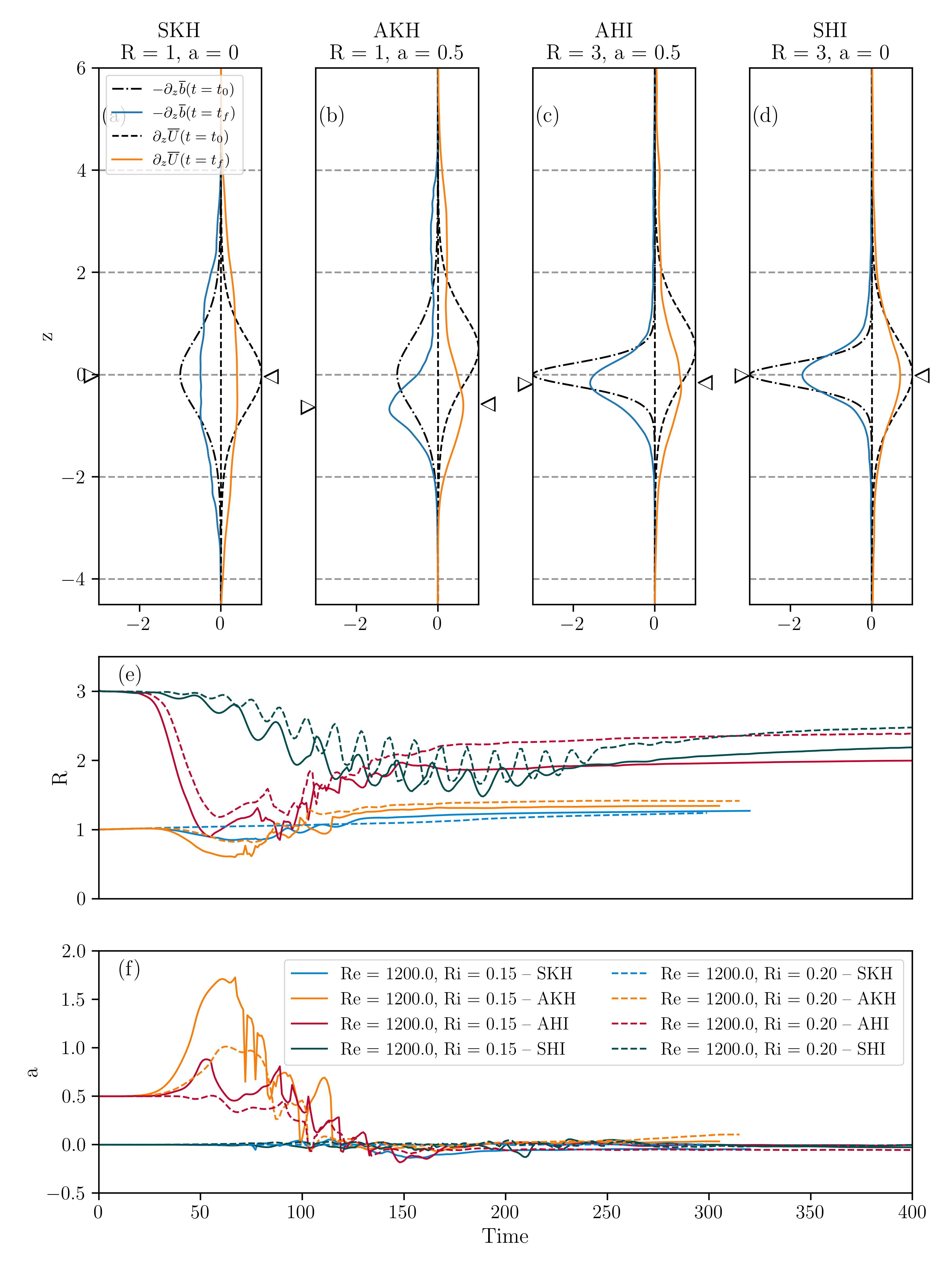}
    \caption{Plot of the initial and final profiles of $-\pderiv{z}{\Havg{b}}$ and $\pderiv{z}{\Havg{u_1}}$ of the (a) SKH, (b) AKH, (c) AHI, and (d) SHI for $\Rey=1200,\ \Ri=0.15$. The evolution of the (e) interface thicknesses $\Rt$ and (f) offset $\at$ are included as a function of time. }
    \label{fig:ParameterEvolve}
\end{figure}

The initial velocity and buoyancy profiles of the SKH, AKH, AHI, and SHI differ in the value of the initial interface thickness ratio $\R$ and the initial interface offset $\offset$. Throughout the mixing event, the mean value of the relative interface thickness and offset evolve in time. To demonstrate this evolution, we have included a comparison of the initial and final $-\pderiv{z}{\Havg{b}}$ and $\pderiv{z}{\Havg{u_1}}$ profiles (taken at the end of each simulation $t=t_f$) in figure~\ref{fig:ParameterEvolve}(a-d). The initially symmetric profiles (SKH, SHI) retain their symmetry. Conversely, initially asymmetric profiles (AKH, AHI) preferentially mix one side of the buoyancy interface more than the other. This asymmetric mixing will displace the position of the buoyancy ($\ZOB$) and velocity ($\ZOU$) interface over the course of the simulation. This is evident by looking at the interface heights at the final time (that is, the heights of the peak gradients in buoyancy and velocity), marked with a $\nabla$ on the corresponding panels of figure~\ref{fig:ParameterEvolve}a-d. Note that the asymmetric mixing results in final profiles in which the peak gradients are approximately aligned, nearly eliminating the initial offset.

In addition to the interface position, we quantify the evolving interface widths as the standard deviation of $\Havg b$ or $\Havg u_1$ about the peak gradient. That is, we define the buoyancy ($\sigma_b$) and velocity ($\sigma_u$) interface thicknesses as 
\begin{gather}
    \sigma_{b} = \sqrt{\frac{\int \left(z-\ZOB\right)^2 \partial_z \overline b \dz}{\int \partial_z \overline b \dz}}, \\
    \sigma_{u} = \sqrt{\frac{\int \left(z-\ZOU\right)^2 \partial_z \overline u_1 \dz}{\int \partial_z \overline u_1 \dz}}.
\end{gather}
Having estimated the interface position and width with time, the evolving interface thickness ratio ($\Rt$) and interface offset ($\at$) are 
\begin{gather}
    \Rt(t) = \frac{\sigma_u}{\sigma_b}, \qquad \at(t) = \ZOU - \ZOB.
\end{gather}
Note that $\Rt(t=0) \approx \R, \ \at(t=0)\approx\offset$. 

We plot the evolution of $\Rt$ in time in figure~\ref{fig:ParameterEvolve}(e) for all $\Rey=1200$ cases, and include the initial and final values of $\Rt$ and $\at$ for all cases in table~\ref{Table:FlowProperties}. We do not include the data from the $\Rey=300$ cases in figure \ref{fig:ParameterEvolve}, as the low Reynolds number results in significant interface diffusion that obfuscates the present results. However, even at $\Rey=300$, we find similar conclusions to those presented here (see table \ref{Table:FlowProperties}).  
%

Several features can be seen in the plot of $\Rt$ (figure~\ref{fig:ParameterEvolve}e). First, the formation of the large billow in the SKH, AKH, and AHI cases is associated with a transient decrease in $\Rt$. On the other hand, a clear oscillatory behaviour is apparent in the SHI flows, consistent with the wavelike character of the flow described above. As the billow or vortices break down into turbulence and the flow mixes, $\Rt$ increases and then nearly plateaus.

The SKH and AKH cases ($\Rt(t=0)=1$) tend towards increased thickness ratios of $\Rt\approx1.3$ by the end of the simulation, consistent with the velocity diffusing faster than the buoyancy for $\Pran=9$. For the instabilities with initially thin buoyancy interfaces ($\Rt(t=0)=3$, i.e.~SHI and AHI), the flows tend towards reduced values of $\Rt\approx2$ by the end of the simulations, indicating that the turbulent event has thickened the buoyancy gradient more than the velocity gradient. However, the buoyancy interface remains thinner than the velocity interface, even with the signature of the initial billow for the AHI case.

Recall that the initial offset between the shear and buoyancy interfaces for the AKH and AHI is nearly eliminated by the end of the mixing events (see figure~\ref{fig:ParameterEvolve}f). The flow asymmetry for these cases results in preferential mixing above the buoyancy interface, leading to a downwards shift of both interfaces. As the shear interface is located above the buoyancy interface, it is mixed at a faster rate, such that the final position of the two interfaces are co-located. The asymmetric mixing reduces the interface offset in the final state.

\begin{table*}
\centering
\begin{tabular}{cc|cc|cc|cc|ccc|ccc}
Case & Instability& $t_{2D,max}$ & $t_{3D,max}$ & $\R$  & $\Rt(t=t_f)$  & $\offset$ & $\at(t=t_f)$& \hspace{1em}$\mathcal{P}$ \hspace{1em} & \hspace{1em} $\mathcal{B}$ \hspace{1em} & \hspace{1em} $\mathcal{D}$ \hspace{1em} & $\eta_M$ & $\overline \eta$& $\overline \eta_{3D}$\\
\hline
1 & SKH & 223.4 &309.3&1.0&1.8&0.0&0.0&             0.11 &  -0.03 & -0.09 &      0.13 & 0.22 & 0.13\\ 
2 & AKH & 175.4 & 235.6 & 1.0 & 1.6 & 0.5 & 0.2 &   0.28 &  -0.07 & -0.21 &      0.16 & 0.60 & 0.20\\ 
3 & AHI & 198.5 & 239.2 & 3.0 &2.2&0.5 &0.0&        0.39 &  -0.08 & -0.31 &     0.17 & 0.48 & 0.28\\ 
4 & SHI & 324.4 & 379.5 & 3.0 & 2.2&0.0&0.0&         0.36 &  -0.08 & -0.27 &      0.17 & 0.62 & 0.23\\ 
5 & SKH & 92.4 & 131.3 & 1.0 & 1.3 & 0.0 &0.0&      0.50 &  -0.10 & -0.40 &     0.21 & 0.36& 0.23\\  
6 & AKH& 60.0 & 134.4 & 1.0 & 1.3 & 0.5 & 0.0&      0.45 &  -0.10 & -0.35 &      0.19 & 0.21& 0.21 \\ 
7 & AHI & 89.6 & 147.9 & 3.0&2.0&0.5&0.0&           0.42 &  -0.07 & -0.35 &     0.17 & 0.23 & 0.22\\ 
8 & SHI & 183.0 & 232.2  & 3.0 & 2.3 & 0.0&0.0&     0.25 &  -0.04 & -0.21 &     0.20 & 0.23 & 0.20\\ 
9 & SKH & 159.3 & 191.6  & 1.0 & 1.2 & 0.0&0.0&     0.04 &  -0.01 & -0.03 &      0.13 & 0.14 & 0.11\\  
10 & AKH & 69.0 & 134.4 & 1.0& 1.4 & 0.5&0.1&       0.25 &  -0.05 & -0.20 &     0.19 & 0.30 & 0.29\\ 
11 & AHI & 128.4 & 160.1 &3.0 & 2.4& 0.5& -0.1&     0.29 &  -0.05 & -0.23 &      0.17 & 0.37 & 0.15\\ 
12 & SHI & 208.0  & 247.4 & 3.0&2.5&0.0&0.0&        0.17 &  -0.03 & -0.14 &     0.20 & 0.31 & 0.20
\end{tabular}
\caption{Table of case parameters for the twelve numerical simulations. Time of maximum 2D and 3D TKE are given as $t_{2D,max}$ and $t_{3D,max}$. The initial and final interface ratios ($\R$, $\Rt(t=t_f)$ ) and interface offsets ($\offset$ , $\at(t=t_f)$) are also included. The time integral of production ($\mathcal{P}$), buoyancy flux ($\mathcal{B}$), and dissipation ($\mathcal{D}$) are included. The median mixing efficiency ($\eta_M$), mean mixing efficiency ($\overline \eta$), and mean mixing efficiency within the turbulent region ($\overline \eta_{3D}$, as described in \S~\ref{sec:SelfOrg}) are given for each case. }
\label{Table:FlowProperties}
\end{table*}

\subsection{Self-organized criticality}\label{sec:SelfOrg}


\begin{figure}
    \centering
    \includegraphics[width=\textwidth]{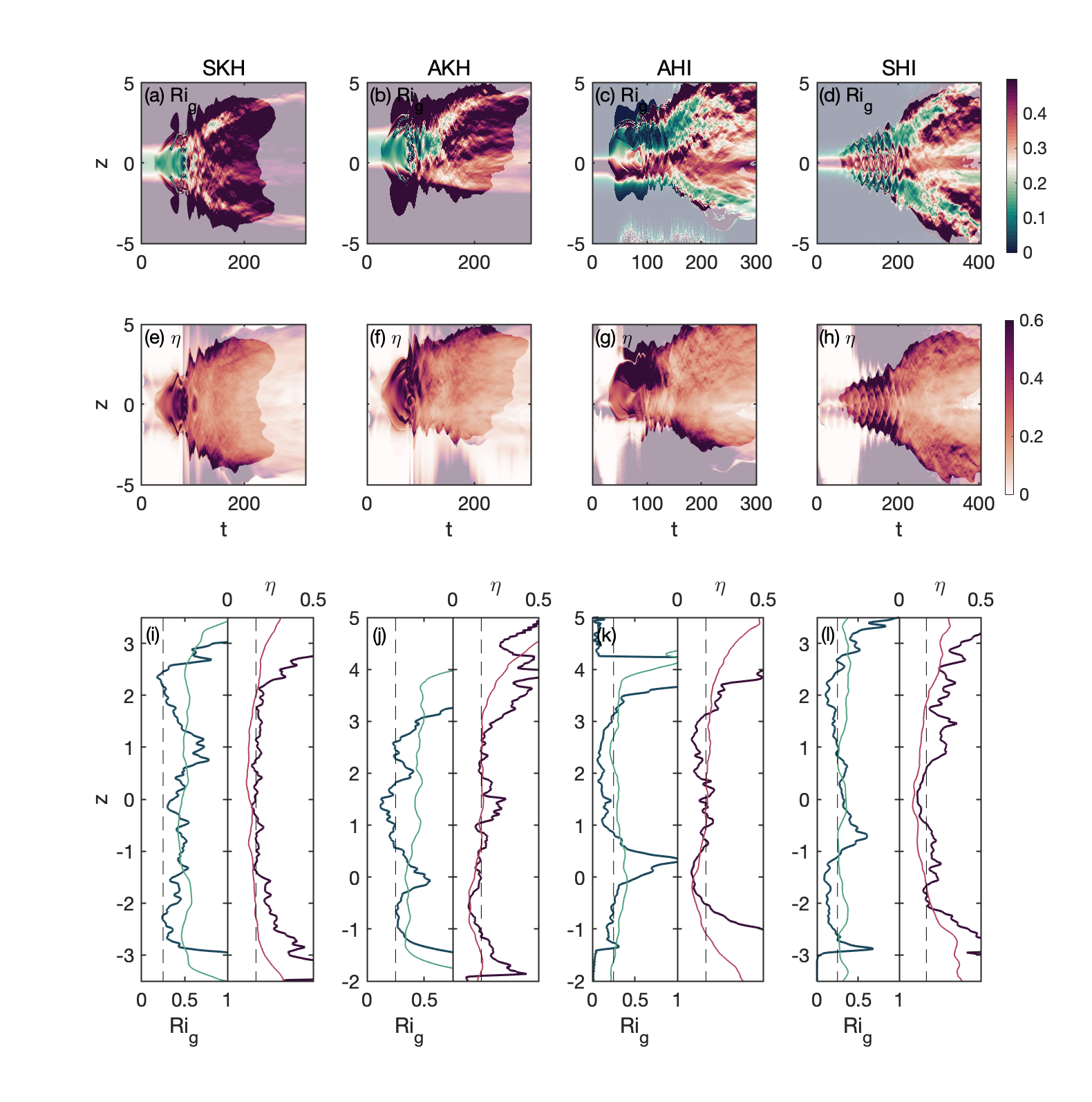}
    \caption{(a-d) Timeseries plots of the gradient Richardson number of the horizontally-averaged flow, $N^2/S^2$. (e-h) Local mixing efficiency, $\eta$, as defined in equation~(\ref{eqn:eta}). The opaque regions in (a-h) indicate the portions of the flow for which $\varepsilon'<2\Ri/(\Rey\ L_z)$. (i-l) Profiles of $Ri_g$ and $\eta$ of the horizontally-averaged flow at $t=t_{3D,max}$ (dark) and averaged from $t_{3D,max}$ (light) to the end of the simulation. The thin dashed lines indicate $Ri_g=1/4$ and $\eta=1/6$. All cases shown have $\Rey=1200$ and $\Ri=0.15$. }
    \label{fig:Rig_eta}
\end{figure}

The gradient Richardson number, 
\begin{gather}\label{eqn::RIG}
    Ri_g = \frac{N^2}{S^2} \, ,
\end{gather}
is an important parameter in studies of stratified shear flows, relating the stabilizing effect of stratification with the destabilizing effect of shear. In particular, the ``critical'' value $Ri_g\sim1/4$ is commonly associated with stratified turbulent mixing: $Ri_g<1/4$ is a necessary condition for instability in steady parallel inviscid stratified shear flows  \citep{Miles,Howard}, and oceanic observations suggest that a critical Richardson number close to $1/4$ is a useful criterion in practice for the onset of turbulent mixing (e.g.~\citealp{Eriksen1978}). Oceanic and estuarine observations suggest that stratified shear flows often exhibit a peak in the distribution of $Ri_g$ around this critical value [e.g.~\citealp{Geyer1987,Smyth2013,Pham2017}]. 

Motivated by these observations, recent studies have focused on the idea of ``self-organized criticality'' in stratified turbulent flows, that is, the tendency of flows to evolve to a state with $Ri_g$ near the critical value. One commonly-invoked argument suggests that this is associated with a balance between external energy input to the background flow (forcing) driving the flow towards a state with $Ri_g<1/4$, allowing for the possibility of shear instability, and overturning turbulent mixing acting to increase $Ri_g$ to values above $1/4$; the balance between these two processes is thought to result in a state of ``marginal stability'' \citep{Smyth2019a,Smyth2020}. 

The above explanation relies on the presence of external forcing to drive the system back to an unstable state. \citet{Salehipour2018} offered an alternate argument for this self-organizing behaviour, namely that symmetric Holmboe instabilities evolve to a state where the average gradient Richardson number is approximately $1/4$, regardless of initial flow parameters. Furthermore, they argue self-organized criticality is not present for KH instability, which can exhibit average gradient Richardson numbers above or below $0.25$, depending on the initial parameters. It should be emphasized that this description is for freely-evolving shear layers like those we consider here, rather than the forced system of \citet{Smyth2019a}.


To what extent do asymmetric shear instabilities share this tendency to \textit{self-organize}?
We have shown that the initial offset in the location of peak buoyancy and shear is essentially removed by the end of the simulation, and that the ratio of interface thicknesses converges to approximately the same values regardless of initial asymmetry. Thus, we may wonder whether there is some ``attractive'' state to which the system is evolving, depending on how Holmboe- or KH-like the initial instability is. 
To explore this question, we investigate the statistics of the gradient Richardson number ($Ri_g$) and the mixing efficiency ($\eta$), defined by equation~(\ref{eqn:eta}), for both the symmetric and asymmetric initial conditions.

We first consider $Ri_g$ in the region where the flow is sufficiently turbulent. As in \citet{Salehipour2018}, we focus on regions where $\varepsilon^\prime>\frac{2}{Lz} \frac{\Ri}{\Rey}$, as this avoids regions where both $N^2,S^2 \to 0$ (see figure \ref{fig:MeanFields}). Similarly, we restrict $t>t_{3D}$, where $t_{3D}$ is the time of maximum kinetic energy, removing the initial laminar period. We note that the results in this paper are relatively robust to variations in the precise values of these two filtering criteria.  

As shown in figure~\ref{fig:Rig_eta}(a,b), the SKH and AKH cases start from an initial state in which $Ri_g$ has a minimum at the centre of the shear layer. After the instability grows and triggers a transition to turbulence, the flow rapidly mixes to a state in which $Ri_g\gtrsim1/4$, though slightly lower values of $Ri_g$ persist for longer in the AKH case than its symmetric counterpart. On the other hand, the AHI and SHI cases, start with a local maximum in $Ri_g$ at the centre of the shear layer (figure~\ref{fig:Rig_eta}c,d). The onset of turbulence in both cases does not lead to the same rapid increase of $Ri_g$. Instead, as a consequence of the turbulence being strongest in the weakly-stratified regions away from the buoyancy interface, the turbulent shear layer shows a range of gradient Richardson numbers for these two cases. 

\begin{figure*}
    \centering
    \includegraphics[width=0.7\textwidth]{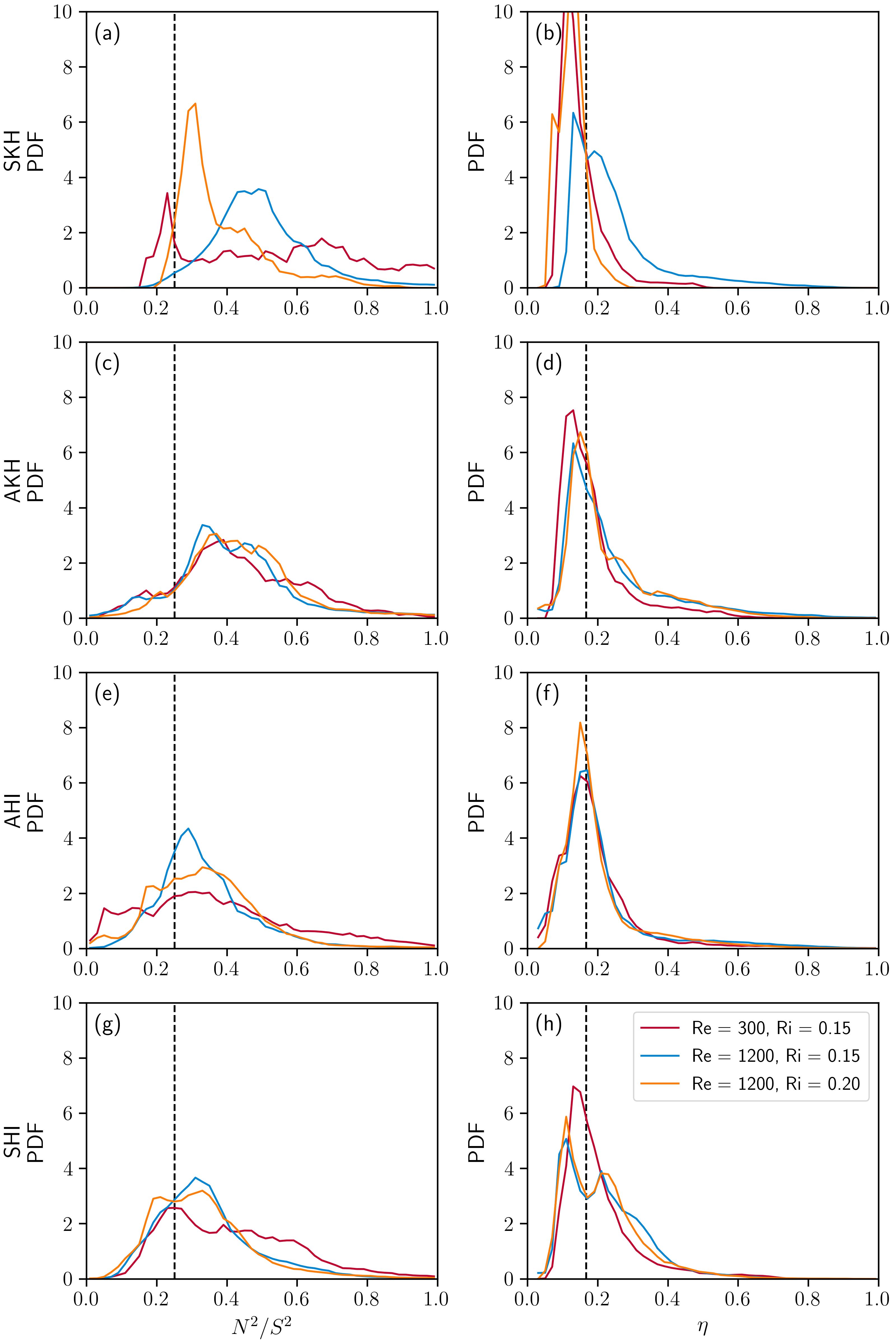}
    \caption{PDFs of $Ri_g$ and $\eta$ where the flow is turbulent for (a,b) SKH, (c,d) AKH, (e,f) AHI, and (g,h) SHI. Vertical dashed lines are included at  $Ri_g=\frac{1}{4}$ and $\eta=\frac{1}{6}$, respectively.}
    \label{fig:Rig_pdf}
\end{figure*}

To quantify these distributions further, within the turbulent region after $t_{3D}$, we compute the probability density function (PDF) of $Ri_g$  (figure~\ref{fig:Rig_pdf}a,c,e,g). Consistent with \citet{Salehipour2018}, the different SKH cases (figure~\ref{fig:Rig_pdf}a) have different peaks in $Ri_g$, and the PDFs of the SHI cases (figure~\ref{fig:Rig_pdf}g) appear to show a consistent peak near $Ri_g=0.25$. 

Of more interest is the behaviour of the AKH and AHI cases (figure~\ref{fig:Rig_pdf}c,e). In contrast to the SKH cases, both asymmetric configurations show a collapse of the PDFs across the parameter values considered here. That is, these PDFs of $Ri_g$ are not very sensitive to the particular values of $\Rey$ and $\Ri$. However, the {\it value} of the peaks in $Ri_g$ differs between the AKH and AHI flows: for AHI, the peak is close to the value of $1/4$ seen in the SHI cases, while AKH show a consistent peak around $Ri_g\approx0.4$ (similar to the larger $Ri_g$ values seen for some of the SKH flows). Again, the stark dependence on $\Rey$ and $\Ri$ associated with SKH appears to be reduced for AKH. However, the details of the turbulent flows still depend on the specific case in question (and the degree to which it is more or less ``KH-like'' versus ``Holmboe-like'').

 We next turn to the corresponding distributions of the mixing efficiency, $\eta$, for the different symmetric and asymmetric cases considered here. The mixing efficiency is the fraction of the total energy loss that increases the potential energy of the system (as opposed to the energy lost to viscous dissipation). 
 Following the approach of \citet{Smith2021}, we define the local mixing efficiency as 
 \begin{gather}
     \eta = \frac{\Ri \chi^\prime}{\Ri \chi^\prime + \varepsilon^\prime} \, , \label{eqn:eta}
 \end{gather}
 which indicates the height where the mixing is most efficient. To avoid regions in which $N^2\rightarrow0$, we apply the same filter as with $Ri_g$.

We show the spatial distribution of $\eta$ in figure~\ref{fig:Rig_eta}(e-h). There is an early peak in $\eta$ in all cases, associated with the initially laminar formation of the billows or counter-propagating vortices. As the flow breaks down into turbulence, the efficiency drops off. For the SKH case, the efficiency is approximately uniform across the turbulent region. Conversely, the AKH, AHI, and SHI cases show a range of values of $\eta$, with lower efficiencies near the interface and larger values away (particularly in the AHI and SHI cases), as a consequence of the larger values of $\chi'$ in those regions. 

The predicted value of $\eta$ for a given flow has been a matter of some debate, as reviewed by \citet{Gregg2018}. In particular, \citet{Osborn1980} suggested an upper bound of 1/6 for the average mixing efficiency of steady homogeneous stratified shear flows. This value, corresponding to a flux coefficient of $\Gamma=\eta/(1-\eta)=0.2$, has been applied in a wide variety of numerical, experimental, and oceanic contexts.

Even for stratified shear layers such as those considered here, the peak value of $\eta$ has been shown to be case specific. For KH instability, $\eta$ has been shown to depend on the flow parameters, the route by which the flow transitions to turbulence, 
and the details of the initial conditions \citep{Caulfield2000,Mashayek2013,Kaminski2019}. On the other hand, studies of scouring-type flows, including Holmboe-driven turbulent mixing \citep{Salehipour2018} and forced stratified shear layers \citep{Smith2021} have shown peaks in the PDF of mixing efficiency near the canonical $1/6$ value, consistent with the flows experiencing a period of steady homogeneous mixing (as assumed by \citet{Osborn1980}). For the simulations presented here, we observe a similar behaviour for the AKH, AHI, and SHI cases, as shown in figure~\ref{fig:Rig_pdf}(d,f,h): the PDFs of $\eta$, defined over the same spatial and temporal region as in the PDFs of $Ri_g$, show peaks around $1/6$ across different flow parameters. On the other hand, the PDFs of $\eta$ for the SKH simulations (figure~\ref{fig:Rig_pdf}b) show peaks at different values depending on the flow parameters, consistent with the results of \citet{Salehipour2018}.

These typical values of $Ri_g$ and $\eta$ can also be seen in the profiles at $t_{3D,max}$ and time-averaged from $t_{3D,max}$ to the end of the simulation in figure~\ref{fig:Rig_eta}(i-l). The SKH and AKH profiles show gradient Richardson numbers above the marginal value of $1/4$, while the AHI and SHI flows show $Ri_g\sim1/4$ over the region surrounding the shear layer. Similarly, local values of $\eta$ are close to $1/6$ during the turbulent phases of the mixing events.

Taken together, the distributions of $Ri_g$ and $\eta$, along with the evolution of the mean quantities described in \S~\ref{sec::MeanQuantities}, suggest that as the underlying instabilities shift from being more KH-like to more Holmboe-like, the corresponding turbulence and mixing likewise transition from one limiting behaviour to the other.

\section{Discussion and conclusions} \label{sec:conclusion}

A substantial fraction of the literature on stratified shear instabilities has focused on the idealized case of a vertically-symmetric background flow, with the centre of the background shear layer coincident with the corresponding buoyancy interface.
However, in many geophysical contexts, the shear and stratification are asymmetric. Even in idealized laboratory experiments, the interfaces may be offset \citep{Lawrence1991,lefauve2018,Yang2022}. We have shown that this asymmetry modifies both the linear stability of the underlying shear instability and its subsequent nonlinear evolution. 

By considering the linear stability of asymmetric flows given by~(\ref{eqn:bgflow_nondim}), we have observed that the eigenfunctions of the unstable modes share characteristics of both pure KH and Holmboe instabilities. Moreover, using an appropriately-defined pseudomomentum \citep[see][]{Eaves2019}, we have quantified how KH-like or Holmboe-like these asymmetric instabilities are for background flows with different $\Ri$, $\Rey$, $\R$, and $\offset$. 

Then, we selected several representative cases ranging from pure KH to pure Holmboe behaviour to demonstrate the effect of asymmetry on the nonlinear evolution of the system. We performed a set of twelve direct numerical simulations with different $\Rey$, $\Ri$, $\offset$ and $\R$. These cases included  symmetric KH [SKH], asymmetric KH [AKH], asymmetric Holmboe [AHI], and symmetric Holmboe [SHI] configurations. These simulations build on the work of \citet{Carpenter2007}, with a higher Reynolds number and the inclusion of the asymmetric KH setup which was not previously considered. From these simulations, and consistent with the linear predictions, we showed that the asymmetric shear instabilities resulted in a flow that had both the propagating wave feature of Holmboe instabilities and the finite-amplitude billow structure of KH instabilities. These hybrid features can be seen in both the TKE budget and the horizontally averaged quantities. As predicted by the pseudomomentum, the nonlinear simulations suggest that there exists a continuous spectrum of KH-like and Holmboe-like behaviour.

As mentioned in the introduction, turbulence caused by stratified shear instabilities is frequently categorized as ``overturning'', in which KH-driven mixing smears out the buoyancy profile, or ``scouring'', where Holmboe-driven mixing preserves the sharp buoyancy interface \citep{Smyth2003,Salehipour2016b,Caulfield2021}. The mixing driven by the asymmetric shear instabilities described here differs from most previous numerical studies in two important ways. 
First, the asymmetric shear instabilities lead to asymmetric mixing of the background buoyancy and momentum -- both the velocity and buoyancy profiles preferentially diffuse on one side of the buoyancy interface. This mixing roughly preserves the relative thickness of the buoyancy and velocity interfaces. Interestingly, in the cases considered, the asymmetric mixing nearly eliminated the initial offset in peak gradients by the end of the simulations. In this sense, the asymmetric mixing produces a more symmetric final state than was present initially. This may have important implications in contexts where the degree of asymmetry changes with time \citep{Yang2022}. Similarly, in a continuously forced system \citep{Smith2021}, it is unclear how the offset would evolve. Second, the mixing appears to share characteristics of both overturning and scouring flows, with part of the initial interface maintained throughout the flow evolution. That is, not only is the mixing vertically asymmetric, but the character of the mixing itself can change depending on the initial offset of the background flow. 

We note that while the simulations presented in this manuscript consider multiple values of $\Rey$ and $\Ri$, further exploration of the parameter space would be of great value in understanding the role of asymmetry in stratified shear flows. In particular, we have kept the Prandtl number fixed at a value of $\Pran=9$ in this study, characteristic of heat in water. The Prandtl number has been shown to play an important role in other studies of shear-driven mixing (e.g.~\citealp{Salehipour2015}) and layer formation (e.g.~\citealp{Taylor2017}); understanding how the Prandtl number may impact the interplay between scouring and overturning discussed in this manuscript may be important to consider for other environmental and geophysical flows. In addition to the Prandtl number, future studies should consider how this spectrum of behaviours changes at higher $\Rey$ more characteristic of environmental flows. Simulations in domains with longer streamwise extent would also be of interest in order to ascertain the importance of upscale cascades of energy via pairing in such asymmetric shear flows [see, e.g.~\citealp{Smyth1991,Mashayek2012a,Dong2019}]. 

In addition to extending the parameter space for flows described by~(\ref{eqn:bgflow_nondim}), other forms of asymmetric stratified shear layers could be considered. For example, \citet{pham_near-n_2012} and \citet{pham_evolution_2014} studied asymmetric flows where shear and stratification varied between the upper and lower layers, and found similar transitions between KH- and Holmboe-like behaviours, similar to the results presented here. Future studies could consider a variety of different types of asymmetry, analyzing both the linear modes using the pseudomomentum framework described here as well as the nonlinear evolution and turbulent mixing. 

It has been suggested that the longer-lived scouring behaviour associated with scouring events could lead to larger overall mixing of the background flow during the turbulent phase compared to the intense burst of mixing associated with KH instability \citep{Smyth2003,Salehipour2016a,Caulfield2021}. It is unclear how this may extend to asymmetric flows: while asymmetry may lead to longer-lived turbulence, the strongest mixing is also offset from the strongest stratification. While it is difficult to draw conclusions at the parameter values studied here, future work with a broader parameter space could explore this question in more detail.   

Oceanographic observations of $Ri_g$ frequently show a peak value of approximately $0.25$ [e.g.~\citealp{Smyth2013,Holleman2016}]. (This value should not be thought of as a result of the Miles-Howard theorem, the underlying assumptions of which do not apply to fully turbulent flows.) One possible explanation for these observations is that forcing and turbulent diffusion act together to drive the flow to a state of marginal stability \citep{Smyth2019a}. 
An alternate explanation is that stratified turbulence driven by Holmboe instability may self-organize into a state with $Ri_g\sim 1/4$ and a flux coefficient $\sim0.2$, behaviour which is not seen for KH-driven mixing \citep{Salehipour2018}. 
Our results are consistent with that conclusion: the SHI cases demonstrate a consistent peak in the PDF of $Ri_g$ near 0.25 that is absent in the SKH cases. Indeed, our SKH simulations include the subcritical, critical, and supercritical cases illustrated in figure~13 of \citet{Salehipour2018}. Of greater interest are the asymmetric cases, which also demonstrate a collapse in the distributions of $Ri_g$, consistent with our interpretation that asymmetry results in mixing events with characteristics of both KH and Holmboe instabilities. 
The AHI has a peak in $Ri_g\approx 0.3$, roughly at the same location as that of the SHI. There is a similar collapse of the AKH curves around Ri$_g\approx$ 0.4. As the SKH cases appear to be highly dependant on the initial flow parameters, this work suggests that even small amounts of asymmetry reduce the dependence of the PDF of $Ri_g$ on the external parameters. Future work aimed at clarifying the role of asymmetry on the resulting distribution of $Ri_g$, particularly at even higher $\Rey$ more typical of geophysical flows and considering a broader range of $\mathcal{R}_M$ to include more instability behaviours, will help to clarify the degree to which the turbulence may self-organize in AKH- and AHI-driven flows. Forced simulations, in which the mechanism described by \citet{Smyth2019a} may also be active, would further allow for the exploration of these self-organizing behaviours in asymmetric stratified shear flows.

We emphasize that the SKH/AKH/AHI/SHI simulations shown in figure~\ref{fig:SliceEvolution} all have the same $\Rey$, $\Pran$, and $\Ri$; the qualitatively different flow evolution arises from small-scale details in the initial shear and stratification profiles. We found a similar range of behaviours in each of the three sets of SKH/AKH/AHI/SHI simulations with fixed $\Rey$, $\Pran$, and $\Ri$.
That is, significantly different behaviours can arise for the same bulk parameters. 
Because of this small-scale dependence, methods to identify the type of instability are of great interest, especially in the analysis of observational data. Given the success of the pseudomomentum approach in classifying the linear dynamics, it is natural to ask whether a similar metric could be found for fully nonlinear flows, either by extending the nonlinear metrics described by \citet{Eaves2019} to less idealized flows or by exploiting newer data-driven analyses for flow classification. 

Stratified turbulent flows may be quantified in terms of a variety of key lengthscales describing, for example, the size of individual overturns or the scales at which stratification becomes important. Understanding the relationships between such lengthscales can be essential for interpreting field measurements in which certain variables are not easily measured. However, most of these relationships for stratified shear instabilities have typically focused on symmetric KH instabilities (e.g.~\citealp{Smyth2000,Mashayek2017}). Future work will consider how these relationships are modified by asymmetry and how they can be related to the specifics of the irreversible mixing \citep{Winters1995}, thereby allowing for improved interpretation of field measurements.

Geophysical flows are rarely perfectly aligned. Our results have highlighted that even a relatively small amount of asymmetry may produce a different flow evolution than classic shear instability theory would suggest. This asymmetry is below the resolvable scales of large-scale circulation models, and as such must be parameterized. In the future, we hope to compare our work with field studies, quantify the importance of asymmetry, and find a practicable solution to incorporate this effect into regional models.

\section{Acknowledgements}

The authors acknowledge helpful discussions with Bill Smyth and Tom Eaves. Jason Olsthoorn was supported by funding from the Natural Sciences and Engineering Research Council of Canada (NSERC) and the Killam Trusts. Alexis Kaminski was supported by National Science Foundation grant OCE-1657676. Daniel Robb was supported by an NSERC doctoral scholarship. The simulations were run through Compute Canada.

\bibliographystyle{abbrvnat}
\bibliography{REF}

\begin{thebibliography}{61}
\providecommand{\natexlab}[1]{#1}
\providecommand{\url}[1]{\texttt{#1}}
\expandafter\ifx\csname urlstyle\endcsname\relax
  \providecommand{\doi}[1]{doi: #1}\else
  \providecommand{\doi}{doi: \begingroup \urlstyle{rm}\Url}\fi

\bibitem[Alexakis(2009)]{Alexakis2009}
A.~Alexakis.
\newblock Stratified shear flow instabilities at large {Richardson} numbers.
\newblock \emph{Phys. Fluids}, 21:\penalty0 054108, 2009.
\newblock \doi{10.1063/1.3147934}.

\bibitem[Brucker and Sarkar(2007)]{Brucker2007}
K.~A. Brucker and S.~Sarkar.
\newblock Evolution of an initially turbulent stratified shear layer.
\newblock \emph{Phys. Fluids}, 19:\penalty0 105105, 2007.
\newblock \doi{10.1063/1.2756581}.

\bibitem[Carpenter et~al.(2007)Carpenter, Lawrence, and Smyth]{Carpenter2007}
J.~R. Carpenter, G.~A. Lawrence, and W.~D. Smyth.
\newblock Evolution and mixing of asymmetric {H}olmboe instabilities.
\newblock \emph{J. Fluid Mech.}, 582:\penalty0 103--132, 2007.
\newblock \doi{10.1017/S002211200700598}.

\bibitem[Carpenter et~al.(2010)Carpenter, Balmforth, and
  Lawrence]{Carpenter2010}
J.~R. Carpenter, N.~J. Balmforth, and G.~A. Lawrence.
\newblock Identifying unstable modes in stratified shear layers.
\newblock \emph{Phys. Fluids}, 22:\penalty0 054104, 2010.
\newblock \doi{10.1063/1.3379845}.

\bibitem[Carpenter et~al.(2011)Carpenter, Tedford, Heifetz, and
  Lawrence]{Carpenter2011}
J.~R. Carpenter, E.~W. Tedford, E.~Heifetz, and G.~A. Lawrence.
\newblock Instability in stratified shear flow: review of a physical
  interpretation based on interacting waves.
\newblock \emph{Appl. Mech. Rev.}, 64:\penalty0 060801, 2011.
\newblock \doi{10.1115/1.4007909}.

\bibitem[Caulfield(2021)]{Caulfield2021}
C.~P. Caulfield.
\newblock Layering, instabilities, and mixing in turbulent stratified flows.
\newblock \emph{Annu. Rev. Fluid Mech.}, 53:\penalty0 113--145, 2021.
\newblock \doi{10.1146/annurev-fluid-042320- 100458}.

\bibitem[Caulfield and Peltier(2000)]{Caulfield2000}
C.~P. Caulfield and W.~R. Peltier.
\newblock The anatomy of the mixing transition in homogeneous and stratified
  free shear layers.
\newblock \emph{J. Fluid Mech.}, 413:\penalty0 1--47, 2000.
\newblock \doi{10.1017/S0022112000008284}.

\bibitem[Caulfield et~al.(1995)Caulfield, Peltier, Yoshida, and
  Ohtani]{Caulfield1995}
C.~P. Caulfield, W.~R. Peltier, S.~Yoshida, and M.~Ohtani.
\newblock An experimental investigation of the instability of a shear flow with
  multilayered density stratification.
\newblock \emph{Phys. Fluids}, 7\penalty0 (12):\penalty0 3028--3041, 1995.
\newblock \doi{10.1063/1.868679}.

\bibitem[Childs et~al.(2012)Childs, Brugger, Whitlock, Meredith, Ahern,
  Pugmire, Biagas, Miller, Harrison, Weber, Krishnan, Fogal, Sanderson, Garth,
  Bethel, Camp, R\"{u}bel, Durant, Favre, and Navr\'{a}til]{VisIt}
H.~Childs, E.~Brugger, B.~Whitlock, J.~Meredith, S.~Ahern, D.~Pugmire,
  K.~Biagas, M.~Miller, C.~Harrison, G.~H. Weber, H.~Krishnan, T.~Fogal,
  A.~Sanderson, C.~Garth, E.~W. Bethel, D.~Camp, O.~R\"{u}bel, M.~Durant, J.~M.
  Favre, and P.~Navr\'{a}til.
\newblock Visit: An end-user tool for visualizing and analyzing very large
  data.
\newblock In \emph{High Performance Visualization--Enabling Extreme-Scale
  Scientific Insight}, pages 357--372. October 2012.
\newblock \doi{10.1201/b12985}.

\bibitem[Corcos and Sherman(1976)]{Corcos1976}
G.~M. Corcos and F.~S. Sherman.
\newblock Vorticity concentration and the dynamics of unstable free shear
  layers.
\newblock \emph{J. Fluid Mech.}, 73\penalty0 (2):\penalty0 241--264, 1976.

\bibitem[Dong et~al.(2019)Dong, Tedford, Rahmani, and Lawrence]{Dong2019}
W.~Dong, E.~W. Tedford, M.~Rahmani, and G.~A. Lawrence.
\newblock Sensitivity of vortex pairing and mixing to initial perturbations in
  stratified shear flows.
\newblock \emph{Phys. Rev. Fluids}, 4:\penalty0 063902, 2019.
\newblock \doi{10.1103/PhysRevFluids.4.063902}.

\bibitem[Eaves and Balmforth(2019)]{Eaves2019}
T.~S. Eaves and N.~J. Balmforth.
\newblock Instability of sheared density interfaces.
\newblock \emph{J. Fluid Mech.}, 860:\penalty0 145--171, 2019.
\newblock \doi{10.1017/jfm.2018.827}.

\bibitem[Eriksen(1978)]{Eriksen1978}
C.~C. Eriksen.
\newblock Measurements and models of fine structure, interval gravity waves,
  and wave breaking in the deep ocean.
\newblock \emph{J. Geophys. Res.}, 83\penalty0 (C6):\penalty0 2989--3009, 1978.

\bibitem[Garrett and Munk(1972)]{Garrett1972}
C.~Garrett and W.~Munk.
\newblock Oceanic mixing by breaking internal waves.
\newblock \emph{Deep-Sea Res.}, 19:\penalty0 823--832, 1972.
\newblock \doi{10.1016/0011-7471(72)90001-0}.

\bibitem[Geyer and Smith(1987)]{Geyer1987}
W.~R. Geyer and J.~D. Smith.
\newblock Shear instability in a highly stratified estuary.
\newblock \emph{J. Phys. Oceanogr.}, 17:\penalty0 1668--1679, 1987.
\newblock \doi{10.1175/1520-0485(1987)017<1668:SIIAHS>2.0.CO;2}.

\bibitem[Gregg et~al.(2018)Gregg, {D'Asaro}, Riley, and Kunze]{Gregg2018}
M.~C. Gregg, E.~A. {D'Asaro}, J.~J. Riley, and E.~Kunze.
\newblock Mixing efficiency in the ocean.
\newblock \emph{Annu. Rev. Mar. Sci.}, 10:\penalty0 443--473, 2018.
\newblock \doi{10.1146/annurev- marine- 121916- 063643}.

\bibitem[Hazel(1972)]{Hazel1972}
P.~Hazel.
\newblock Numerical studies of the stability of inviscid stratified shear
  flows.
\newblock \emph{J. Fluid Mech.}, 51\penalty0 (1):\penalty0 39--61, 1972.

\bibitem[Holleman et~al.(2016)Holleman, Geyer, and Ralston]{Holleman2016}
R.~C. Holleman, W.~R. Geyer, and D.~K. Ralston.
\newblock Stratified turbulence and mixing efficiency in a salt wedge estuary.
\newblock \emph{J. Phys. Oceanogr.}, 46:\penalty0 1769--1783, 2016.
\newblock \doi{10.1175/JPO-D-15-0193.1}.

\bibitem[Holmboe(1962)]{Holmboe1962}
J.~Holmboe.
\newblock On the behavior of symmetric waves in stratified shear layers.
\newblock \emph{Geofys. Publ.}, 24:\penalty0 67--113, 1962.

\bibitem[Howard(1961)]{Howard}
L.~N. Howard.
\newblock Note on a paper of john w. miles.
\newblock \emph{Journal of Fluid Mechanics}, 10\penalty0 (4):\penalty0
  509–512, 1961.
\newblock \doi{10.1017/S0022112061000317}.

\bibitem[Kaminski and Smyth(2019)]{Kaminski2019}
A.~K. Kaminski and W.~D. Smyth.
\newblock Stratified shear instability in a field of pre-existing turbulence.
\newblock \emph{Journal of Fluid Mechanics}, 862:\penalty0 639–658, 2019.
\newblock \doi{10.1017/jfm.2018.973}.

\bibitem[Kaminski et~al.(2021)Kaminski, {D'Asaro}, Shcherbina, and
  Harcourt]{Kaminski2021}
A.~K. Kaminski, E.~A. {D'Asaro}, A.~Y. Shcherbina, and R.~R. Harcourt.
\newblock High-resolution observations of the {North Pacific} transition layer
  from a {Lagrangian} float.
\newblock \emph{J. Phys. Oceanogr.}, 51:\penalty0 3163--3181, 2021.
\newblock \doi{10.1175/JPO-D-21-0032.1}.

\bibitem[Klaassen and Peltier(1991)]{Klaassen1991}
G.~P. Klaassen and W.~R. Peltier.
\newblock The influence of stratification on secondary instability in free
  shear layers.
\newblock \emph{J. Fluid Mech.}, 227:\penalty0 71--106, 1991.

\bibitem[Lawrence et~al.(1991)Lawrence, Browand, and Redekopp]{Lawrence1991}
G.~A. Lawrence, F.~K. Browand, and L.~G. Redekopp.
\newblock The stability of a sheared density interface.
\newblock \emph{Phys. Fluids A}, 3\penalty0 (10):\penalty0 2360--2370, 1991.
\newblock \doi{10.1063/1.858175}.

\bibitem[Lefauve et~al.(2018)Lefauve, Partridge, Zhou, Dalziel, Caulfield, and
  Linden]{lefauve2018}
A.~Lefauve, J.~L. Partridge, Q.~Zhou, S.~B. Dalziel, C.~P. Caulfield, and P.~F.
  Linden.
\newblock The structure and origin of confined holmboe waves.
\newblock \emph{Journal of Fluid Mechanics}, 848:\penalty0 508–544, 2018.
\newblock \doi{10.1017/jfm.2018.324}.

\bibitem[Lewin and Caulfield(2021)]{Lewin2021}
S.~F. Lewin and C.~P. Caulfield.
\newblock The influence of far-field stratification on shear-induced turbulent
  mixing.
\newblock \emph{J. Fluid Mech.}, 928:\penalty0 A20, 2021.
\newblock \doi{10.1017/jfm.2021.755}.

\bibitem[Mashayek and Peltier(2012{\natexlab{a}})]{Mashayek2012a}
A.~Mashayek and W.~R. Peltier.
\newblock The `zoo' of secondary instabilities precursory to stratified shear
  flow transition. part 1 {Shear} aligned convection, pairing, and braid
  instabilities.
\newblock \emph{J. Fluid Mech.}, 708:\penalty0 5--44, 2012{\natexlab{a}}.
\newblock \doi{10.1017/jfm.2012.304}.

\bibitem[Mashayek and Peltier(2012{\natexlab{b}})]{Mashayek2012b}
A.~Mashayek and W.~R. Peltier.
\newblock The `zoo' of secondary instabilities precursory to stratified shear
  flow transition. part 2 {The} influence of stratification.
\newblock \emph{J. Fluid Mech.}, 708:\penalty0 45--70, 2012{\natexlab{b}}.
\newblock \doi{10.1017/jfm.2012.294}.

\bibitem[Mashayek and Peltier(2013)]{Mashayek2013}
A.~Mashayek and W.~R. Peltier.
\newblock Shear-induced mixing in geophysical flows: does the route to
  turbulence matter to its efficiency?
\newblock \emph{J. Fluid Mech.}, 725:\penalty0 216--261, 2013.
\newblock \doi{10.1017/jfm.2013.176}.

\bibitem[Mashayek et~al.(2017)Mashayek, Salehipour, Bouffard, Caulfield,
  Ferrari, Nikurashin, Peltier, and Smyth]{Mashayek2017}
A.~Mashayek, H.~Salehipour, D.~Bouffard, C.~P. Caulfield, R.~Ferrari,
  M.~Nikurashin, W.~R. Peltier, and W.~D. Smyth.
\newblock Efficiency of turbulent mixing in the abyssal ocean circulation.
\newblock \emph{Geophys. Res. Lett.}, 44:\penalty0 6296--6306, 2017.
\newblock \doi{10.1002/2016GL072452}.

\bibitem[Miles(1961)]{Miles}
J.~W. Miles.
\newblock On the stability of heterogeneous shear flows.
\newblock \emph{Journal of Fluid Mechanics}, 10\penalty0 (4):\penalty0
  496–508, 1961.
\newblock \doi{10.1017/S0022112061000305}.

\bibitem[Moum et~al.(2003)Moum, Farmer, Smyth, Armi, and Vagle]{Moum2003}
J.~N. Moum, D.~M. Farmer, W.~D. Smyth, L.~Armi, and S.~Vagle.
\newblock Structure and generation of turbulence at interfaces strained by
  internal solitary waves propagating shoreward over the continental shelf.
\newblock \emph{J. Phys. Oceanogr.}, 33:\penalty0 2093--2112, 2003.

\bibitem[Moum et~al.(2011)Moum, Nash, and Smyth]{Moum2011}
J.~N. Moum, J.~D. Nash, and W.~D. Smyth.
\newblock Narrowband oscillations in the upper equatorial ocean. part {I}:
  interpretation as shear instabilities.
\newblock \emph{J. Phys. Oceanogr.}, 41:\penalty0 397--411, 2011.
\newblock \doi{10.1175/2010JPO4450.1}.

\bibitem[Osborn(1980)]{Osborn1980}
T.~R. Osborn.
\newblock Estimates of the local rate of vertical diffusion from dissipation
  measurements.
\newblock \emph{Journal of Physical Oceanography}, 10\penalty0 (1):\penalty0 83
  -- 89, 1980.
\newblock \doi{10.1175/1520-0485(1980)010<0083:EOTLRO>2.0.CO;2}.

\bibitem[Pawlak and Armi(1998)]{Pawlak1998}
G.~Pawlak and L.~Armi.
\newblock Vortex dynamics in a spatially accelerating shear layer.
\newblock \emph{J. Fluid Mech.}, 376:\penalty0 1--35, 1998.

\bibitem[Pham and Sarkar(2014)]{pham_evolution_2014}
H.~T. Pham and S.~Sarkar.
\newblock Evolution of an asymmetric turbulent shear layer in a thermocline.
\newblock \emph{Journal of Turbulence}, 15\penalty0 (7):\penalty0 449--471,
  July 2014.
\newblock ISSN 1468-5248.
\newblock \doi{10.1080/14685248.2014.914216}.
\newblock URL
  \url{http://www.tandfonline.com/doi/abs/10.1080/14685248.2014.914216}.

\bibitem[Pham et~al.(2012)Pham, Sarkar, and Winters]{pham_near-n_2012}
H.~T. Pham, S.~Sarkar, and K.~B. Winters.
\newblock Near-{N} {Oscillations} and {Deep}-{Cycle} {Turbulence} in an
  {Upper}-{Equatorial} {Undercurrent} {Model}.
\newblock \emph{Journal of Physical Oceanography}, 42\penalty0 (12):\penalty0
  2169--2184, Dec. 2012.
\newblock ISSN 0022-3670, 1520-0485.
\newblock \doi{10.1175/JPO-D-11-0233.1}.
\newblock URL
  \url{https://journals.ametsoc.org/view/journals/phoc/42/12/jpo-d-11-0233.1.xml}.
\newblock Publisher: American Meteorological Society Section: Journal of
  Physical Oceanography.

\bibitem[Pham et~al.(2017)Pham, Smyth, Sarkar, and Moum]{Pham2017}
H.~T. Pham, W.~D. Smyth, S.~Sarkar, and J.~N. Moum.
\newblock Seasonality of deep cyycle turbulence in the eastern equatorial
  {P}acific.
\newblock \emph{J. Phys. Oceanogr.}, 47:\penalty0 2189--2209, 2017.

\bibitem[Salehipour et~al.(2015)Salehipour, Peltier, and
  Mashayek]{Salehipour2015}
H.~Salehipour, W.~R. Peltier, and A.~Mashayek.
\newblock Turbulent diapycnal mixing in stratified shear flows: the influence
  of {Prandtl} number on mixing efficiency and transition at high {Reynolds}
  number.
\newblock \emph{J. Fluid Mech.}, 773:\penalty0 178--223, 2015.
\newblock \doi{10.1017/jfm.2015.225}.

\bibitem[Salehipour et~al.(2016{\natexlab{a}})Salehipour, Caulfield, and
  Peltier]{Salehipour2016b}
H.~Salehipour, C.~P. Caulfield, and W.~R. Peltier.
\newblock Turbulent mixing due to the {Holmboe} wave instability at high
  {Reynolds} number.
\newblock \emph{J. Fluid Mech.}, 803:\penalty0 591--621, 2016{\natexlab{a}}.
\newblock \doi{10.1017/jfm.2016.488}.

\bibitem[Salehipour et~al.(2016{\natexlab{b}})Salehipour, Peltier, Whalen, and
  {MacKinnon}]{Salehipour2016a}
H.~Salehipour, W.~R. Peltier, C.~B. Whalen, and J.~A. {MacKinnon}.
\newblock A new characterization of the turbulent diapycnal diffusivities of
  mass and momentum in the ocean.
\newblock \emph{Geophys. Res. Lett.}, 43:\penalty0 3370--3379,
  2016{\natexlab{b}}.
\newblock \doi{10.1002/2016GL068184}.

\bibitem[Salehipour et~al.(2018)Salehipour, Peltier, and
  Caulfield]{Salehipour2018}
H.~Salehipour, W.~R. Peltier, and C.~P. Caulfield.
\newblock Self-organized criticality of turbulence in strongly stratified
  mixing layers.
\newblock \emph{Journal of Fluid Mechanics}, 856:\penalty0 228–256, 2018.
\newblock \doi{10.1017/jfm.2018.695}.

\bibitem[Smith et~al.(2021)Smith, Caulfield, and Taylor]{Smith2021}
K.~M. Smith, C.~Caulfield, and J.~Taylor.
\newblock Turbulence in forced stratified shear flows.
\newblock \emph{Journal of Fluid Mechanics}, 910:\penalty0 A42, 2021.
\newblock \doi{10.1017/jfm.2020.994}.

\bibitem[Smyth(2020)]{Smyth2020}
W.~D. Smyth.
\newblock Marginal instability and the efficiency of ocean mixing.
\newblock \emph{J. Phys. Oceanogr.}, 50:\penalty0 2141--2150, 2020.
\newblock \doi{10.1175/JPO-D-20-0083.1}.

\bibitem[Smyth and Moum(2000)]{Smyth2000}
W.~D. Smyth and J.~N. Moum.
\newblock Length scales of turbulence in stably stratified mixing layers.
\newblock \emph{Physics of Fluids}, 12\penalty0 (6):\penalty0 1327--1342, 2000.
\newblock \doi{10.1063/1.870385}.

\bibitem[Smyth and Moum(2013)]{Smyth2013}
W.~D. Smyth and J.~N. Moum.
\newblock Marginal instability and deep cycle turbulence in the eastern
  equatorial pacific ocean.
\newblock \emph{Geophysical Research Letters}, 40\penalty0 (23):\penalty0
  6181--6185, 2013.
\newblock \doi{https://doi.org/10.1002/2013GL058403}.

\bibitem[Smyth and Peltier(1991)]{Smyth1991}
W.~D. Smyth and W.~R. Peltier.
\newblock Instability and transition in finite-amplitude {Kelvin-Helmholtz} and
  {Holmboe} waves.
\newblock \emph{J. Fluid Mech.}, 228:\penalty0 387--415, 1991.

\bibitem[Smyth and Winters(2003)]{Smyth2003}
W.~D. Smyth and K.~B. Winters.
\newblock Turbulence and mixing in {Holmboe} waves.
\newblock \emph{J. Phys. Oceanogr.}, 33:\penalty0 694--711, 2003.

\bibitem[Smyth et~al.(2011)Smyth, Moum, and Nash]{Smyth2011}
W.~D. Smyth, J.~N. Moum, and J.~D. Nash.
\newblock Narrowband oscillations in the upper equatorial ocean. {Part II}:
  {P}roperties of shear instabilities.
\newblock \emph{J. Phys. Oceanogr.}, 41:\penalty0 412--428, 2011.
\newblock \doi{10.1175/2010JPO4451.1}.

\bibitem[Smyth et~al.(2019)Smyth, Nash, and Moum]{Smyth2019a}
W.~D. Smyth, J.~D. Nash, and J.~N. Moum.
\newblock Self-organized criticality in geophysical turbulence.
\newblock \emph{Scientific Reports}, 9:\penalty0 3747, 2019.
\newblock \doi{10.1038/s41598-019-39869-w}.

\bibitem[Strang and Fernando(2001)]{Strang2001}
E.~J. Strang and H.~J.~S. Fernando.
\newblock Entrainment and mixing in stratified shear flows.
\newblock \emph{J. Fluid Mech.}, 428:\penalty0 349--386, 2001.

\bibitem[Subich et~al.(2013)Subich, Lamb, and Stastna]{Subich}
C.~J. Subich, K.~G. Lamb, and M.~Stastna.
\newblock Simulation of the navier–stokes equations in three dimensions with
  a spectral collocation method.
\newblock \emph{International Journal for Numerical Methods in Fluids},
  73\penalty0 (2):\penalty0 103--129, 2013.
\newblock ISSN 1097-0363.
\newblock \doi{10.1002/fld.3788}.

\bibitem[Taylor and Zhou(2017)]{Taylor2017}
J.~R. Taylor and Q.~Zhou.
\newblock A multi-parameter criterion for layer formation in a stratified shear
  flow using sorted buoyancy coordinates.
\newblock \emph{J. Fluid Mech.}, 823:\penalty0 R5, 2017.
\newblock \doi{10.1017/jfm.2017.375}.

\bibitem[Tedford et~al.(2009)Tedford, Carpenter, Pawlowicz, Pieters, and
  Lawrence]{Tedford2009}
E.~W. Tedford, J.~R. Carpenter, R.~Pawlowicz, R.~Pieters, and G.~A. Lawrence.
\newblock Observation and analysis of shear instability in the fraser river
  estuary.
\newblock \emph{Journal of Geophysical Research: Oceans}, 114\penalty0 (C11),
  2009.
\newblock \doi{https://doi.org/10.1029/2009JC005313}.

\bibitem[Thorpe(2018)]{Thorpe2018}
S.~A. Thorpe.
\newblock Models of energy loss from internal waves breaking in the ocean.
\newblock \emph{J. Fluid Mech.}, 836:\penalty0 72--116, 2018.
\newblock \doi{10.1017/jfm.2017.780}.

\bibitem[Tu et~al.(2020)Tu, Fan, Lian, Liu, Liu, Kaminski, and Smyth]{Tu2020}
J.~Tu, D.~Fan, Q.~Lian, Z.~Liu, W.~Liu, A.~Kaminski, and W.~Smyth.
\newblock Acoustic observations of {Kelvin-Helmholtz} billows on an estuarine
  lutocline.
\newblock \emph{J. Geophys. Res. Oceans}, 125:\penalty0 e2019JC015383, 2020.
\newblock \doi{10.1029/2019JC015383}.

\bibitem[{van Haren} et~al.(2014){van Haren}, Gostiaux, Morozov, and
  Tarakanov]{vanHaren2014}
H.~{van Haren}, L.~Gostiaux, E.~Morozov, and R.~Tarakanov.
\newblock Extremely long {Kelvin-Helmholtz} billow trains in the {Romanche
  Fracture Zone}.
\newblock \emph{Geophys. Res. Lett.}, 41:\penalty0 8445--8451, 2014.
\newblock \doi{10.1002/2014GL062421}.

\bibitem[{VanDine} et~al.(2021){VanDine}, Pham, and Sarkar]{VanDine2021}
A.~{VanDine}, H.~T. Pham, and S.~Sarkar.
\newblock Turbulent shear layers in a uniformly stratified background: {DNS} at
  high {Reynolds} number.
\newblock \emph{J. Fluid Mech.}, 916:\penalty0 A42, 2021.
\newblock \doi{10.1017/jfm.2021.212}.

\bibitem[Winters et~al.(1995)Winters, Lombard, Riley, and D'Asaro]{Winters1995}
K.~B. Winters, P.~N. Lombard, J.~J. Riley, and E.~A. D'Asaro.
\newblock Available potential energy and mixing in density-stratified fluids.
\newblock \emph{Journal of Fluid Mechanics}, 289:\penalty0 115--128, 1995.
\newblock ISSN 0022-1120.
\newblock \doi{10.1017/S002211209500125X}.

\bibitem[Woods et~al.(2010)Woods, Caulfield, Landel, and Kuesters]{Woods2010}
A.~W. Woods, C.~P. Caulfield, J.~R. Landel, and A.~Kuesters.
\newblock Non-invasive turbulent mixing across a density interface in a
  turbulent {Taylor-Couette} flow.
\newblock \emph{J. Fluid Mech.}, 663:\penalty0 347--357, 2010.
\newblock \doi{10.1017/S0022112010004295}.

\bibitem[Yang et~al.(2022)Yang, Tedford, Olsthoorn, and Lawrence]{Yang2022}
A.~J.~K. Yang, E.~W. Tedford, J.~Olsthoorn, and G.~A. Lawrence.
\newblock Asymmetric holmboe instabilities in arrested salt-wedge flows.
\newblock \emph{Physics of Fluids}, 34\penalty0 (3):\penalty0 036601, 2022.
\newblock \doi{10.1063/5.0083765}.
\newblock URL \url{https://doi.org/10.1063/5.0083765}.

\end{thebibliography}

\end{document}